# Quantized Acoustic Phonons Map the Dynamics of a Single Virus

Yaqing Zhang[±], Rihan Wu[±], Md Shahjahan[±], Canchai Yang[**], Dohun Pyeon[**], Elad Harel[±*]

[±]Department of Chemistry, Michigan State University, 578 South Shaw Lane, East Lansing MI 48824

[**]Department of Microbiology, Genetics, and Immunology, Michigan State University, 567 Wilson Road, East Lansing, MI 48824

**Corresponding Author**

*elharel@msu.edu

**Abstract:** The natural vibrational frequencies of biological particles such as viruses and bacteria encode critical information about their mechanical and biological states as they interact with their local environment and undergo structural evolution. However, detecting and tracking these vibrations within a biological context at the single particle level has remained elusive. In this study, we track the vibrational motions of single, unlabeled virus particles under ambient conditions using ultrafast spectroscopy. The ultrasonic spectrum of an 80-100 nm lentiviral pseudovirus reveals vibrational modes in the 19-22 GHz range sensitive to virus morphology and 2-10 GHz modes with nanosecond dephasing times reflecting viral envelope protein interactions. By tracking virus trajectories over minutes, we observe acoustic mode coupling mediated by the local environment. Single particle tracking allows capture of viral disassembly through correlated mode softening and dephasing. The sensitivity, high resolution, and speed of this approach promise deeper insights into biological dynamics and early-stage diagnostics at the single microorganism level.

The low-frequency vibrations of biological systems such as proteins, viruses, and bacteria, arise from collective motion of all their constituent atoms. The vibrational spectra of these biological systems, therefore, reflect their three-dimensional structure and conformational flexibility, as well as critical interactions with their environment. However, detecting these low-frequency vibrations which occur in the hundreds of MHz to THz range within a biological environment has remained out of reach(*1*). Current methods capable of detecting narrow spectral regions of these low-frequency vibrations rely on external devices such as ultrahigh frequency (UHF) mechanical resonators(*2*, *3*) or plasmonic nanostructures(*4*) that are specifically tuned to the analyte of interest. However, these platforms are incompatible within a biological environment. Further, these methods restrict the motion of the particles, and, therefore, do not allow tracking of dynamics as the particle environment changes. While studies using low-frequency Raman spectroscopy have been carried out on highly concentrated virus suspensions(*5*, *6*), these ensemble measurements resulted in broad, poorly resolvable spectra. On the other hand, theoretical calculations using an atomistic approach(*7*, *8*) suggested that virus identification should be possible using low-frequency Raman scattering if sufficient resolution is achieved.

    In this study, we demonstrate an *all-optical*, ultra-high resolution, method for detection and tracking of quantized acoustic vibrations in a small, biological particle – a single virus smaller than 100 nm. We measure acoustic spectra in the 2 - 50 GHz range with sub-GHz resolution that are especially sensitive to morphology and interactions of the viral envelope proteins with the

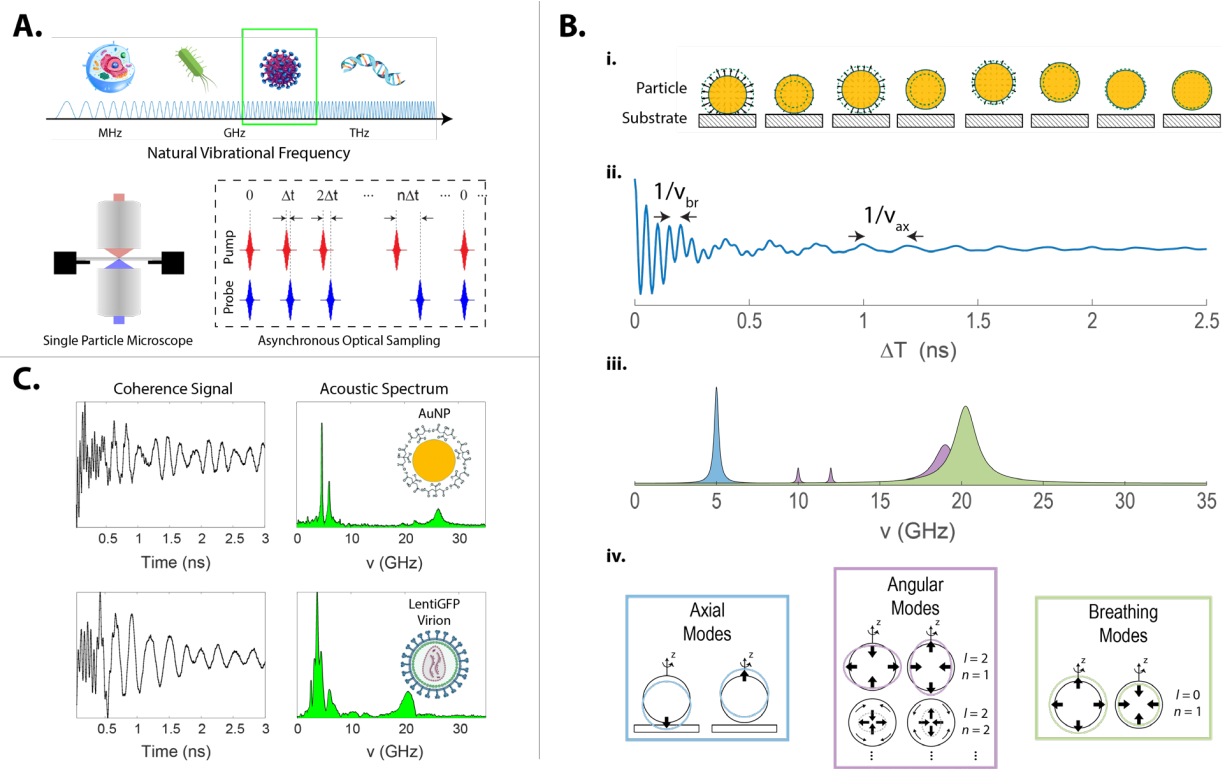

**Figure 1. Principle of *BioSonic* spectroscopy**. **A.** Natural frequencies of biological systems at different length scales. Simplified microscope setup showing illumination and detection geometry. The asynchronous optical sampling (ASOPS) sequence uses a fixed frequency offset to generate a rapid, linear scanning of the time delays up to the laser pulse period. **B.** (i) Motion of a nanoparticle (NP) on a substrate. (ii) Time-domain response showing simultaneous excitation of breathing (br), angular, and axial (ax) modes in the NP. (iii) Corresponding spectrum by Fourier transformation of the time-domain signal. (iv). Different types of vibrational modes that the NP may experience. **C.** Representative time response and spectra of a single ~100 nm gold particle (AuNP), top, and a single lentivirus particle, bottom. Inset shows the VSV-G glycoprotein and membrane envelope (blue) and capsid containing the GFP gene and protein (maroon).

environment. The concept of this approach, which we have termed *BioSonic* spectroscopy, is shown in Figure 1. The virus sample is placed on a cover slip without further modification and interrogated with a pair of ultrashort laser pulses inside a microscope: a non-resonant pump pulse (<100 fs, 1040 nm) to excite collective vibrations in a single virus particle, and a second, time-delayed probe pulse (<100 fs, 785 nm) to detect changes in light scattering induced by the coherent vibrations. The weak signal is isolated from the large background of backscattered light by using balanced detection and asynchronous optical sampling (ASOPS)(*9, 10*), a method by which the inter-pulse delays are rapidly scanned up to the laser pulse period (~10 ns) in sub-milliseconds to reduce laser and environmental noise. Further details of the optical setup and the signal isolation procedures are provided in the supplementary information (SI, (**I., II.**)).

The time-domain signal includes various contributions depending on the resonance condition. The coherence-only (i.e. oscillatory) component of the signal reflects the acoustic phonon spectrum, composed of low-frequency Raman-active vibrational modes in the 0.1 GHz – 2 THz spectral region. Measurement of the pump-probe response provides information on both the acoustic frequency, $v$, as well as the phonon dephasing time, $\Gamma$. This dephasing is dictated by an interplay between the intrinsic anharmonicity of the lattice and other extrinsic factors such as material impurities or defects, and interactions with the local environment. Depending on the particle size, shape, and composition, three different types of vibrational modes are typically observed (Figure 1B): 1) an axial or contact mode(*11*) at low frequency (< 10 GHz), which represents interactions of the particle with the local environment, 2) non-spherically symmetric angular modes (10 – 20 GHz), which correspond to higher order vibrational motion represented by spherical harmonics, and 3) a breathing or radial mode (>20 GHz), which represents radially symmetric vibrational motion(*12*). Angular modes may also be induced by breaking of the particle spherical symmetry near the substrate or from coupling of modes due to the local environment as will be discussed later. For comparison, we show the spectra (Figure 1C) of a ~100 nm spherical gold nanoparticle (AuNP) and an 80-100 nm lentiviral pseudovirus particle with a green fluorescence protein (GFP) gene inserted into its RNA genome (LentiGFP). For these particles, both spectra exhibit fast oscillations that persist for ~500 ps and slower oscillations that persist for at least 5 ns. These oscillations correspond primarily to the breathing and axial modes, respectively, with the virus particle spectra exhibiting a more complex structure at frequencies below 10 GHz.

To better understand the complex acoustic spectrum of the virus, we first investigated a similarly sized spherical AuNP under identical experimental conditions. The properties of the prominent radial breathing mode have been extensively studied using time-resolved ultrafast spectroscopy in a wide range of metallic nanoparticles including gold, silver, and bi-metallics, and in varying sizes and shapes (*12–14*). When metallic NPs are illuminated with an ultrashort pulse, electrons within the Fermi energy are excited to higher lying states in the conduction band forming hot carriers which rapidly thermalize in tens of femtoseconds. This is followed by thermalization with the lattice through electron-phonon interactions, which generates a photo-induced stress in the NP, launching coherent mechanical vibrations that are detected by a time-delayed probe pulse. Above a few nanometers, the acoustic properties of these systems are well-described by an elastic continuum model(*15*), which predicts that the radial breathing frequency scales inversely with the characteristic dimension of the nanostructure and with the square root of the ratio of the shearing modulus, $G$, to the particle density, $\rho$: $v_{br} \propto D^{-1}(G/\rho)^{1/2}$, where D is the particle diameter. For a free particle with radius, $R$, using stress-free boundary conditions (free sphere model (FSM); see SI (**XIII.**)), the energy eigenvalues may be obtained for Lamb's equation of motion for a three-dimensional elastic body(*16, 17*). These eigenvalues depend on the orbital angular momentum quantum number $l$, and harmonic $n$ (see Figure 1B). Since the excitation process occurs through Raman scattering, selection rules dictate that for a free spherical particle only the spherical modes are allowed: $l = 0$ corresponds to a purely radial mode with spherical symmetry, while $l = 2$ is a quadrupolar mode. When the NP is in contact with a surface, in addition to the radial breathing mode, a low-frequency axial mode is observed which depends on the adhesion force between the NP and the local environment. This interaction gives rise to a periodic motion of the particle position relative to the environment (e.g. substrate) which manifests as a few GHz mode for ~100 nm AuNPs. Studies of the breathing and axial modes for different diameter spherical AuNPs have shown that they are related to one another according to classical Hertzian contact mechanics, whereby $v_{ax} \propto v_{br}^{7/6}$, where $v_{br}$ is the radial breathing mode frequency and $v_{ax}$ is the axial mode frequency (*11*, *18*). Therefore, the scaling between the two modes is independent of the particle size.

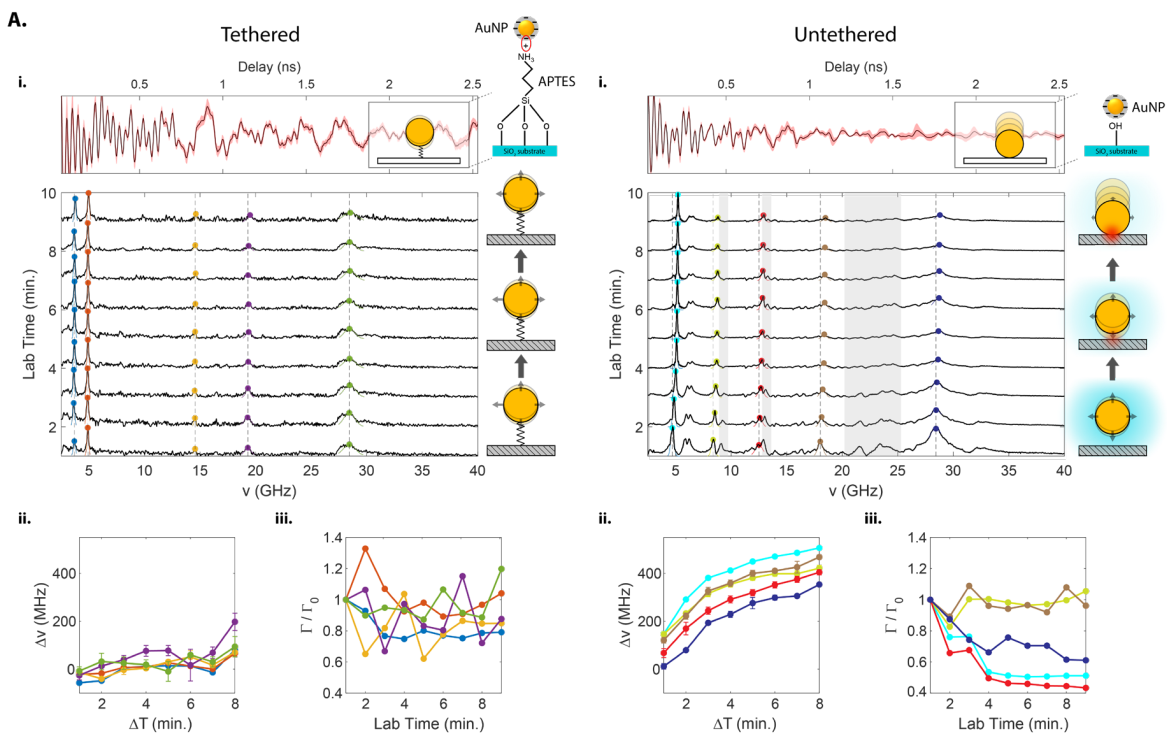

**Figure 2. Tracking of acoustic spectra for tethered (A) from untethered (B) gold nanoparticles (AuNPs).** (i) Top: Average time-domain response (black curve) and standard deviation (filled red curve) over the trajectory. Inset shows the AuNP interaction with the substrate. Bottom: acoustic spectra recorded in 1-minute increments as the laser irradiation modifies the matrix (blue region in (B)) environment. Note, in the tethered case, the environment is considered 'matrix-free'. (ii) Frequency peak shift relative to first time trace for peaks color-coded in (i). (iii). Relative peak broadening as a function of lab time. Note, the spectra are normalized to the highest peak in the spectral window. $v$: frequency, $\Gamma$: dephasing rate. Gray regions show modes that are due to NP/matrix interactions.

Figure 2A(i) shows the acoustic response from a single AuNP that is electrostatically bound to the substrate by a 0.5 – 1 nm long– a (3-Aminopropyl)triethoxysilane (APTES) molecule absorbed on the -OH terminated fused silica substrate(*19*). The molecular tether ensured weak coupling between the AuNP and the substrate and established a more stable and 'matrix-free', environment under laser illumination. The sample preparation procedure (see SI (**IV.**)) ensures minimal residual contaminants and organic impurities. The measured radial breathing mode at $v = 28.92$ GHz agrees with the FSM model for a 105 nm diameter spherical AuNP. At low frequency, the spectrum shows two axial modes near 3.65 GHz and 4.70 GHz: the splitting may be explained by considering the tether acting like a mass on a spring in series between the NP and the substrate. We also observed an angular mode near 18 GHz region which is in close agreement to $l = 2, n = 2$ angular mode expected by FSM.

Next, we considered an 'untethered' AuNP where the adhesion forces between the NP and substrate are much stronger due to direct contact (Figure 2B). Unlike the tethered case, the drop cast procedure resulted in the presence of residual contaminants on the surface which we loosely refer to as the *matrix*, giving rise to variance in the acoustic spectra of individual AuNPs due to heterogeneity in the local environment. For the particles shown in Figure 2, the average frequency of the breathing mode is similar (< 0.35 GHz across the time series), implying that the characteristic diameters are within 1 nm of one another. For the untethered AuNP, the amplitudes of the axial and angular modes were much stronger relative to the breathing mode than those for the tethered particle. We also observed additional acoustic modes for the untethered particle due to breaking of the spherical symmetry near the substrate. The large number of modes observed (within 6 – 25 GHz, Figure 2B) cannot be accounted for by considering only the FSM, which suggests strong coupling between the axial, angular, and breathing modes induced by the local environment. In a spherical particle, there is an $2l + 1$ degeneracy from the z-component of the angular momentum. However, deviation from spherical symmetry give rise to a splitting into $l + 1$ modes(*17, 20*).

To gain further insight into the origin of the substrate-induced matrix coupling, we performed dynamic single-particle spectroscopy (d-SPS) experiments. The lasers, in addition to executing the pump-probe experiments, served to alter the matrix of the NP and the resulting spectral evolution revealed correlations and couplings among the observed acoustic modes. For the tethered AuNP, tracking the particle over 10 minutes revealed only a minor shift (<200 MHz) of the axial, angular, and breathing modes in the 2 – 30 GHz range, as well as no discernable correlation among the mode frequencies (Fig. 2A(ii./iii.)) due to the more stable environment. In the untethered AuNP, however, evolution of the spectral features over several minutes revealed correlations among the mode frequencies as shown in Figure 2B (ii/iii). For the five modes analyzed, selected due to their relative isolation from nearby peaks, the acoustic frequencies all blue-shifted (shift to higher energy) with experimental time. This blueshift occurred to varying degrees for many of the single particle measurements performed (see SI (**VI.**), for other single particle trajectories). The laser light slowly ablated or otherwise removed the matrix surrounding the NP, which both lowered the effective mass and caused stronger association between the NP and the substrate. Such correlations are ideally measured by d-SPS because variations in size, shape, and other material properties such as defects, multiple crystal facets and dislocations, and ligand coverage obscure spectral changes originating from changes in the NP-environment interactions(*21*). As shown in Figure 2B(ii), the blueshift of the axial mode (labelled in cyan) is most pronounced, while the radial breathing mode shift (labelled in blue) is smallest both in absolute ($\Delta v$) and relative terms ($\Delta v/v$). We also observed that the breathing, axial, and one of the angular modes (labelled in red) narrowed over the 10-minute trajectory (Figure 2B(iii)). Additionally, we observed that the amplitudes of the axial mode, relative to the breathing and angular modes, increase over the time series (note, the plots are shown on a normalized scale for clarity). The axial mode amplitude depends on the distance between the NP and substrate as well as the displacement of the breathing mode. d-SPS provides a means to distinguish modes that arise from NP-matrix interaction from those due to NP-substrate interactions. As the matrix is removed and the particle-substrate distance decreases, the breathing mode induces a larger axial amplitude vibration. For the main axial mode and the angular mode in red (12.5 – 12.9 GHz), the dephasing rate decreased (i.e. exhibited a longer lifetime) with increased NP/substrate interaction, which is commensurate with an increase in the breathing mode lifetime (blue marker). Therefore, damping of these surface-sensitive modes are strongly affected by the matrix environment. In contrast, for the two modes shown in green (8.3 – 8.8 GHz) and brown (18 – 18.5 GHz), we observed that the frequency shifts were nearly identical with time. The dephasing of these two modes also followed the same trend, whereby the dephasing rates remained largely unchanged. This suggest that these modes are associated with internal modes (e.g., the $l=2, n=2$ mode shown in Figure 1B), where the surrounding matrix has minimal effect on damping. In some of the broader modes (gray shaded regions), the amplitudes rapidly decrease with experiment time, indicating that these modes arise primarily from NP/matrix interactions(*22*).

With a deeper understanding of how the different acoustic modes in the AuNP influence one another and their dependence on the particle/environment interaction, we turned our attention to the virus particles. In contrast to the metal NPs, the virus vibrations are excited under non-resonant conditions (electronic transitions of biomolecules composing a virus range from about 4.4 – 4.8 eV; compared to 1.2 eV pump energy). Therefore, the excitation proceeds by stimulated Raman scattering(*23*), whereby Raman-active normal modes are excited on the electronic ground state. The signal strength is far weaker than a comparatively sized metal NP (see SI(**III.**) for comparison), and an isolated virus particle does not experience an appreciable temperature change during this process.

Many single particle trajectories were examined (see SI(**VIII.**) for other trajectories), but here we focus on two that clearly illustrate the environmental effect on the virus acoustic spectra. Figure 3A shows the dynamic tracking of a virus particle (Virus #1) that is well isolated and positioned atop a bare fused silica substrate over a span of 12 minutes (see SI(**VII.**) for correlated atomic force microscopy (AFM) image). The notable absence of axial modes suggests that the virus particle was weakly or non-interacting with its environment. The spectra reveal a single breathing mode near 21.8 GHz, with a broad peak at 1-2 GHz lower energy. As with the NPs, basic features of the acoustic spectra may be described by FSM. For the lowest-order radial breathing mode ($l = 0, n = 1$) observed at ~21.8 GHz, the estimated particle diameter is 83 nm using assumptions of the longitudinal sound velocity in the virus (*24*), which agrees with the estimated diameter of 80 – 100 nm for the LentiGFP virus. The next lowest allowed mode at $l = 2, n = 2$ for a spherical, model virus occurs at 20.5 GHz, which is in close agreement with the observed mode at $T \geq 1$ min near 20.8 GHz (Fig. 3A). We also note that slight deviations of the virus shape from nominally spherical may also induce asymmetric line shapes and additional features near the main breathing mode(*15, 25*). The spectra remain largely unchanged until about 6 minutes when the breathing mode begins to weaken and broaden. The mode exhibits a dramatic red shift and acute broadening at 10 minutes before the signal disappears below the noise floor. This red shift and broadening indicate a softening of the phonon modes which occurs as the viral capsid begins to weaken, swell, and suddenly ruptures(26). Assuming the radial breathing mode may be modeled by an underdamped harmonic oscillator (see SI(**XII.**)), the virus initially has a stiffness of $k = 6490 \pm 62 \text{ N/m}$, which then suddenly drops at a rate of $-608 \text{ N/m/min}$ before rupturing (last three time points in Figure 3E(i)). While most virus particles measured stayed intact during the measurement and despite employing non-resonant excitation conditions, the virus particles occasionally ruptured after several minutes of laser exposure, possibly through multi-photon ionization or localized and excess thermal accumulation. Further insight may be gained by examining the evolution of the breathing mode (Figure 3E(ii)), where we observe that the frequency and dephasing rate are anti-correlated with each other (open blue circles and open red triangles). This indicated that weakening of the capsid caused rapid dephasing as the symmetry was reduced and the surface area increased.

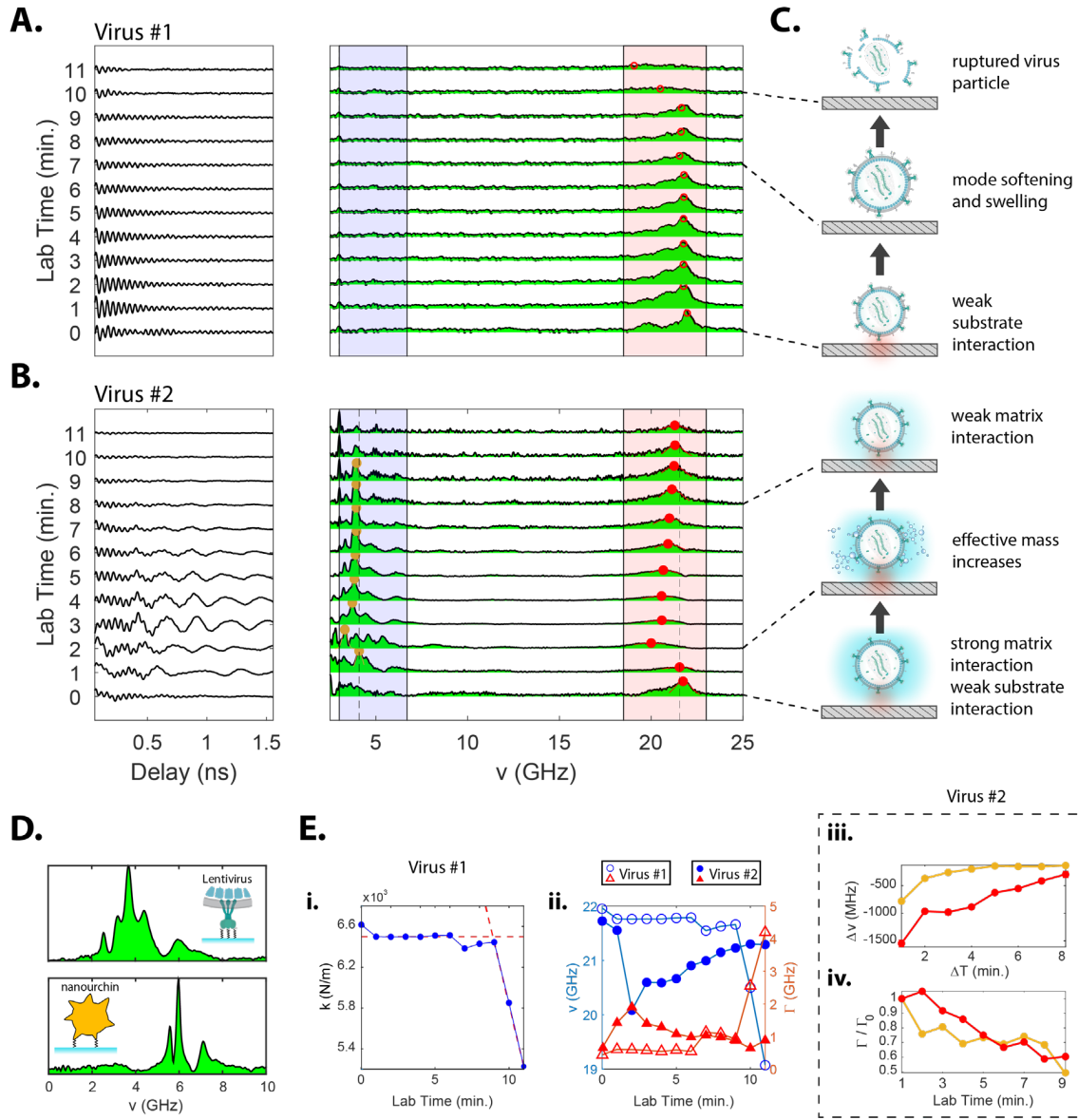

**Figure 3. Single-particle trajectories provide insight into the virus-substrate interactions. A.** Virus #1 trajectory over 12 minutes showing gradual phonon softening and sudden virus particle rupturing. **B.** Virus #2 trajectory showing both axial and breathing modes as well as weaker angular modes. **C.** Cartoon of the state of the virus particle at select time points. **D.** Top: zoomed in spectrum of Virus #2 at T = 3 min in the 0 -10 GHz region showing the axial modes formed from interaction of the glycoproteins with the surrounding matrix. Bottom: zoomed in spectrum of nanourchin particle showing axial mode splitting. **E.** (i) Estimated stiffness of the Virus #1 with lab time (horizontal red dashed line is average over first 9 min.; slanted red dashed line is a fit over the last three minutes). (ii) Correlations between the breathing mode frequency and dephasing rates (closed triangles and circles are for Virus#2; open markers are for Virus #1. (iii). Frequency shift of the breathing and axial modes referenced to the T = 1 min. time trace. (iv). Change in the relative dephasing rate for the breathing mode peak (red) and axial mode peak (orange) for Virus #2 as a function of time.

We then measured a second virus particle (Figure 3B), with the breathing mode and shoulder very similar to Virus #1 initially, but with a much stronger axial mode that changes over time. For this sample, labeled as Virus #2, correlative AFM imaging shows that the particle is buried in a matrix environment (see AFM images in the SI(**VII.**)). The localized environment of each virus particle measured may be different due to material remaining after purification – buffer, impurities, and crystallization of paraformaldehyde which is used for virus deactivation. After 1 minute, the breathing mode in Virus #2 red shifted by 1.5 GHz, broadening significantly. This large red shift of the breathing mode is due to changes in the virus particle/matrix interaction upon repeated laser irradiation. While challenging to characterize, the red shift suggested an increase in the effective mass of the virus, which could be due to some matrix material adhering to it temporally. At 2 minutes, a prominent axial peak became evident (orange circle near 2 GHz region) which started to blueshift as the interaction between the virus particle and matrix decreased. Meanwhile, both the breathing and axial modes experienced a blueshift along with line shape narrowing, corresponding to increasing dephasing time. Angular modes appeared in the 5-10 GHz region, but were weak and difficult to quantify, so they were not evaluated further. The signal starts to weaken at about 8 minutes and never fully recovers to their original state indicating that the virus particle has undergone some damage during the tracking experiment. Other trajectories showed that the virus remained intact for long periods or exhibited partially reversible dynamics in the limit of weak virus/environment interactions (see SI(**VIII./IX.**)).

The axial modes show a far more complex pattern of peaks than in the case of spherical AuNPs (Figure 3D, Top). The high resolution of the acoustic spectrum in the sub-10 GHz spectral region arises from the long lifetime of the axial modes. The features may be qualitatively rationalized by considering the structure of the envelope proteins. The virus particle has the vesicular stomatitis virus G protein (VSV-G, 58.4 kDa without glycosylation) composing the envelope. These glycoproteins may be considered as a series of coupled oscillators whose interactions with the local environment are reflected in mode splitting. In the case of the tethered AuNP, we observed that the axial modes arising from splitting due to the tether acting as an additional mass on a spring. While examining the detailed interactions of the viral envelope proteins with the matrix is beyond the scope of the current study, we considered a biomimetic virus-like NP(*27*) - a single Au 'nanourchin' particle(*28*) which has spiky protrusions coming out of its surface (see SI(**XI.**) for additional details). As shown in figure 3D (bottom), the asymmetric shape of the nanourchin leads to a broad breathing mode, while exhibiting axial mode splitting analogous to those observed in Virus #2.

To gain more insights into the coupling of these modes, we looked at correlations among the spectral features. As with the trajectory of virus #1, the spectral shift of the breathing mode in virus #2 was anti-correlated to the dephasing rate. We then identified which axial modes most strongly coupled to the breathing modes. For the peak marked in orange (2-3 GHz), both the axial and breathing mode increased their lifetime with lab time at nearly the same rate, while both experienced correlated spectral shifts at different relative magnitudes (Figure 3E (iii/iv)). This indicates that the breathing and axial modes are affected by the matrix environment in a correlated manner. The axial modes here are likely less influenced directly by the fused silica substrate as in the case of the AuNP because of the weaker van Der Waals interactions between the virus envelope proteins and the untreated glass. To test this idea, we performed experiments with the same conditions of single virus particles on a mica substrate (grade V-1) where the interaction with envelope proteins is expected to be stronger. We observed an ~3 GHz red shift (see SI(**X.**)) of the breathing mode peak compared to fused silica, which is due to the increase in the effective mass of the particle from the stronger van Der Waals interactions.

We then performed a statistical analysis on 12 individual single virus particle trajectories measured in this study (see correlative AFM in the SI(**VII.**)). The AFM imaging showed a large variance in the environment of each virus particle, which is reflected in a commensurate variance in the measured acoustic spectra. This is attributed to strong environmental effects due to coupling of the axial, angular, and breathing modes as described in detail above. To analyze the large volume of data acquired, we performed a Bayesian analysis method (*29*) on each trajectory that extracted the frequency, dephasing/decay rate, phase, and amplitude of more than 200 individual acoustic spectra (see SI(**XIV.**) for details). Figure 4 shows a 2D histogram and its projection of all the frequencies and decay rates extracted from the statistical analysis. Colored regions in the 2D plot indicate the number of counts within the specified 2D bin region. The most common mode observed is the radial breathing mode near 22 GHz with a dephasing rate of ~1 GHz (1 ns lifetime). We also see the red-shifted mode appear in most of the traces although its position relative to the zero-order breathing mode varies by ~1 GHz. We observe relatively few instances of modes in the 5-15 GHz range, where we would expect other spheroidal modes (besides $l = 0, n = 1$) presumably because these modes are weak and more difficult to identify in the analysis. In the axial mode region, we observe a wide range of frequencies in the 2-5 GHz region with a wide range of dephasing times. We note that the extreme sensitivity of the axial mode to the virus/environment interactions causes these modes to shift appreciably, thereby giving rise to a far wider distribution of frequencies and dephasing rates compared to the relatively stable breathing mode. Ensemble measurements average out over these strong local effects, leading to broad and featureless spectra. This underscores the need to make single particle measurements with high time resolution.

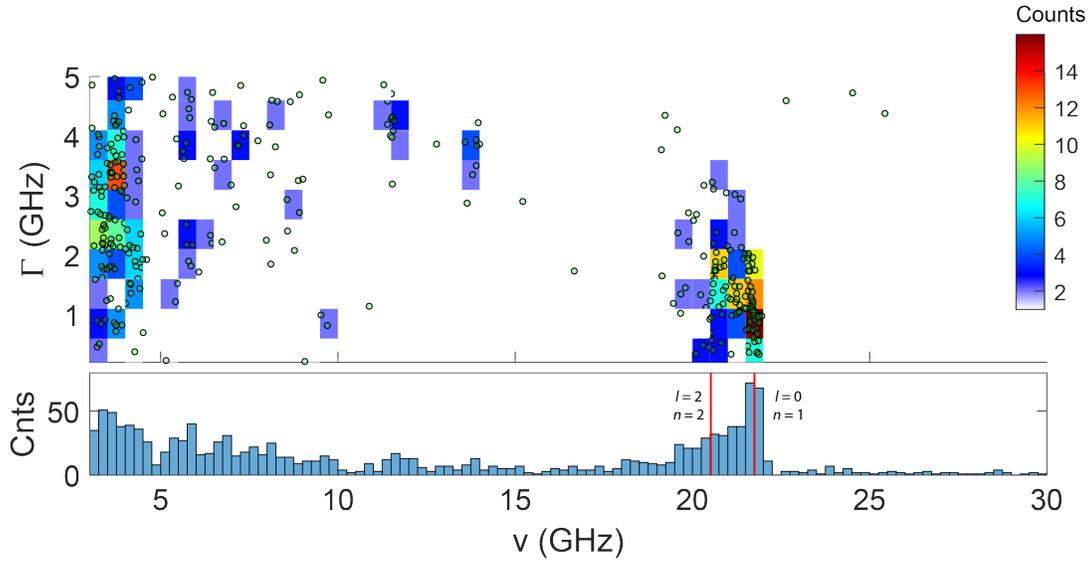

**Figure 4. Statistical analysis highlights the matrix effect.** Summary depiction of all frequencies and decay rates extracted from the statistical analysis. Top: Two-dimensional histogram resulting from analysis of all frequencies recovered by Bayesian analysis. The dephasing rate, $\Gamma$, and frequency, v, are binned by 0.5 GHz x 0.25 GHz. The color bar shows the number of counts per bin. Bottom: The projection of the 2D histogram onto the frequency dimension shows the main breathing mode near 22 GHz and lower-frequency axial and angular modes. Red vertical lines show predicted acoustic modes based on Lamb's model for a free spherical particle.

In conclusion, we have measured the acoustic phonon spectra of individual virus particles by ultrafast spectroscopy for the first time. In contrast to optical spectroscopic measurements of local bond vibrations(*30*), the acoustic spectra are a measure of the collective oscillations of all the atoms that compose the virus particle. The nanosecond lifetimes of these collective vibrations impart to them a remarkable sensitivity to the particle shape and morphology as well as to the interactions of its envelope proteins. Furthermore, tracking the evolution of the acoustic spectra provides unique insight into the effect of the surrounding environment on the vibrational motion(*31*). The method described has sufficient time resolution to examine a single virus particle through its life cycle(*32*) which occurs on the second to hour time scale. Future studies will focus on assigning specific features in the virus particle acoustic spectrum through examination of structure and matrix effects by correlative electron microscopy and other single particle characterization methods(*33*). Investigations on the effect of damping of acoustic vibrations in a liquid environment(*34–36*) will be important for assessing the viability of this approach for *in vivo* applications. While the studies here were focused on viruses, the broad spectral range of *BioSonic* spectroscopy could capture other microorganisms including bacteria and fungi, smaller nanoscale molecular machines such as molecular motors, and large proteins. The ability to detect single, unlabeled virus particles without physical contact enables a long list of applications from ultra-sensitive viral detection to fundamental studies of viral dynamics including self-assembly and infection, paving the way for a comprehensive understanding of many biological processes by correlating both static structures and dynamics.

**Acknowledgments:** This work was supported by the W.M. Keck Foundation and the Defense Threat Reduction Agency (HDTRA12310028).

**Supplementary Information**

**Quantized Acoustic Phonons Map the Dynamics of a Single Virus**

Yaqing Zhang[±], Rihan Wu[±], Md Shahjahan[±], Canchai Yang[**], Dohun Pyeon[**], Elad Harel[±*]

[±]Department of Chemistry, Michigan State University, 578 South Shaw Lane, East Lansing MI 48824

[**]Department of Microbiology, Genetics, and Immunology, Michigan State University, 567 Wilson Road, East Lansing, MI 48824

**Corresponding Author**

*elharel@msu.edu

# Table of Contents

**(I) Optical Setup:** The experimental setup is shown in Figure S1. It consists of an asynchronous optical sampling (ASOPS) laser system, an optical trigger generation system, a home-built correlative microscope, balanced detector, detection electronics, and high-speed digitizer. These sub-systems are described in detail below:

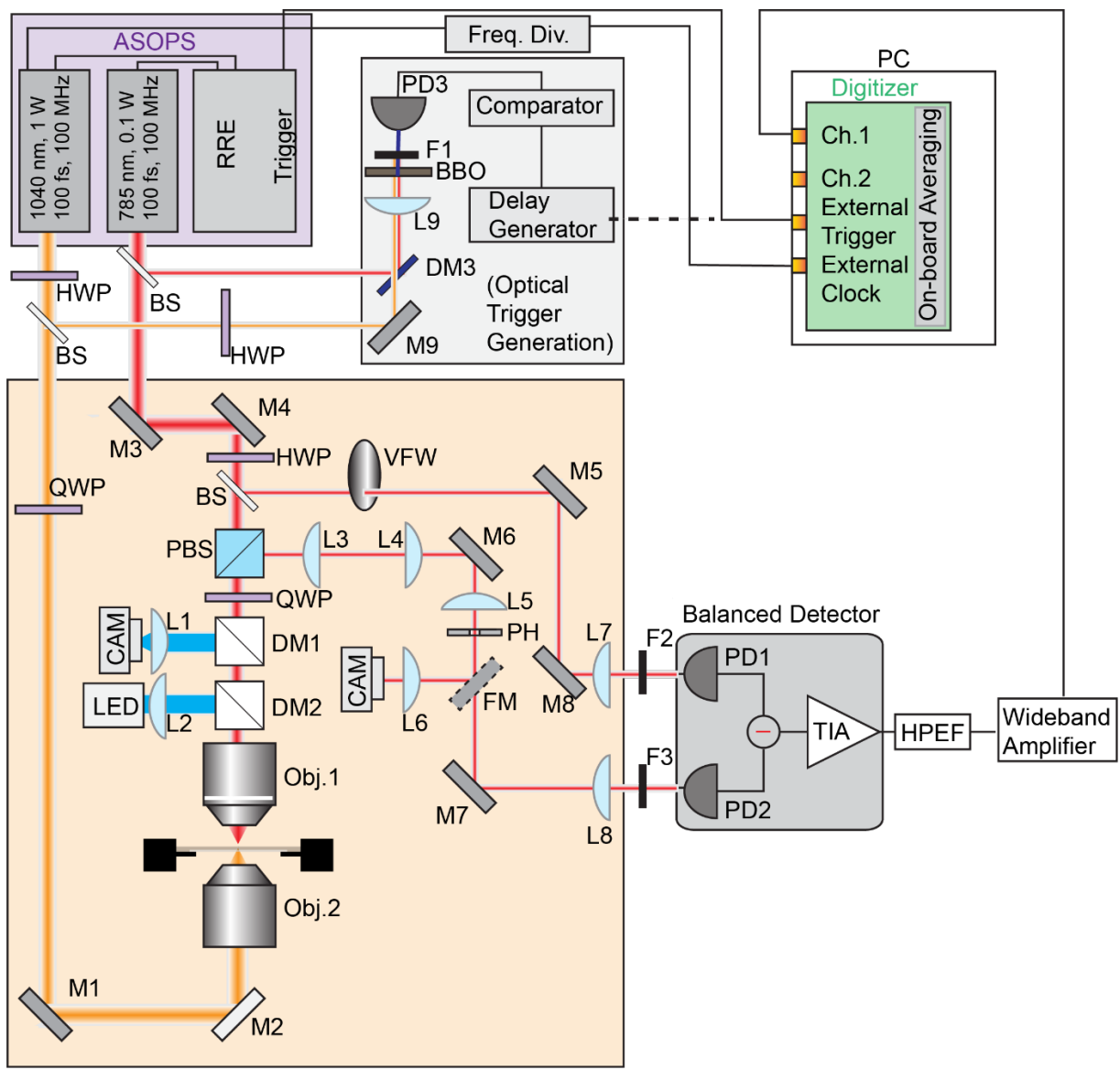

**Figure S1.** Experimental Setup. Asynchronous optical sampling (ASOPS) generates two pulse trains with a fixed repetition rate offset. The pump (orange) and probe pulse (red) are directed towards the sample from opposite directions. The probe pulse is split prior to the sample and serves as a reference for balanced detection. The probe light scattered from the sample is collected in an epi configuration and directed towards one channel of the balanced detector. The signal is then amplified and sent to a high-speed digitizer whose clock is set by the repetition rate electronics (RRE) and triggered by the offset either generated from the RRE electronically or through an optical trigger generation system. HWP = half wave plate, QWP = quarter wave plate, BS = beam splitter, PBS = polarizing beam splitter, DM = dichroic mirror, Obj = objective lens, FM = flip mirror, VFW = variable filter wheel, PH = pinhole, PD = photodiode, BBO = beta barium oxide. HPEF = high pass electrical filter, TIA = transimpedance amplifier. M = mirror, L = lens, F = filter, Freq. Div. = frequency divider.

a. **ASOPS:** The asynchronous optical sampling laser system (ASOPS, Menlo Systems) consists of three components: the pump laser, the probe laser, and the repetition rate electronics (RRE). The pump laser beam is delivered by the ASOPS with a central wavelength of 1040 nm, output power of 1 W, pulse duration of 100 fs, and a repetition rate $f_r = 100$ MHz. The probe laser has a central wavelength of 785 nm, 100 mW output power, 100 fs pulse duration, and a repetition rate $f_r \pm \Delta f$, where $\Delta f$ is a small offset (typically 1-10 kHz). The RRE allows for synchronization between the lasers.

*Scanning Principle:* With ASOPS, the pulse trains scan in time at a rate given by $\Delta f$ (Figure S2). The offset value determines the scan period, $1/\Delta f$, while the maximum measurement window is the pulse train interval, $T_{max} = 1/f_r$. At 100 MHz repetition rate, the measurement window is 10 ns, giving an instrument-limited Fourier resolution of 0.1 GHz. The time delay in the molecular time frame is the difference between the pulse periods of the two laser pulse trains $1/(f_r - \Delta f) - 1/f_r$, which is simplified as $\Delta f/f_r^2$ under the assumption that $\Delta f \ll f_r$. Since the scan is linear and periodic, a GHz to THz frequency of a molecular vibration is down converted to a kHz to MHz digitization frequency in lab time. The down-conversion factor is $\alpha = \Delta f/f_r$. For example, if the offset is 1 kHz and the measurement window is fixed as 10 ns, the scanning period is 1 ms in real time, the scanning resolution is 100 fs, then the down-conversion factor is 1 kHz/100 MHz = $1\times10^{-5}$. In this case, a 1 THz vibrational frequency is down-converted to 10 MHz, while a 1 GHz vibration is down-converted to 10 kHz.

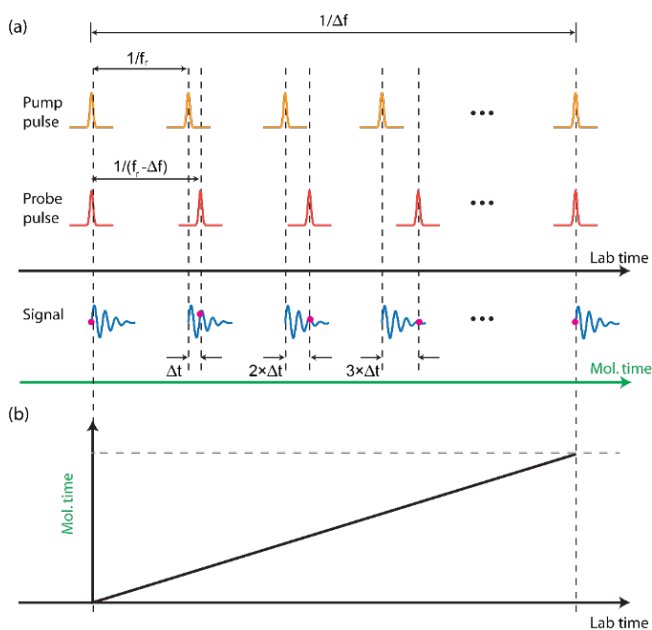

**Figure S2.** Scanning principle of ASOPS. (a) Offset frequency $\Delta f$ determines the scan rate between the two pulse trains. The signal is down-converted from molecular time to lab time by the ratio of the offset to the laser repetition rate. (b) Relation between lab time and molecular time.

b. **Signal Detection:** The signal detection system consists of a balanced detector and electronic signal conditioning:

*Balanced Detection:* The 785 nm probe laser beam is routed by a mirror (M7), and focused to one of the receiving channels of a balanced, amplified detector (PDB210A, Thorlabs). The other channel of the balanced detector receives a reference optical signal, which is achieved by splitting ~10% of the probe beam before it reaches the microscope, using a plate beam splitter (BSN10R, Thorlabs). A set of short pass filters (labelled as F2-3, FESH0900, Thorlabs) and iris apertures are utilized to block the pump beam and stray light, respectively. Note that a combination of fixed neutral density (ND) filters and a variable ND filter wheel (labelled as VFW, NDM2, Thorlabs), allows is for control of the power in the reference arm. After balancing of the two channels, the difference current is converted to a voltage signal and amplified by an internal transimpedance amplifier (TIA) inside the balanced detector. The transimpedance gain is ~ $10^5$ V/A.

*Electronic Signal Conditioning:* The output analog signal from the balanced detector undergoes the following signal conditioning steps:

1) A DC block electric filter (EF599, Thorlabs) and a high pass electric filter (designated as HPEF, e.g. EF115, Thorlabs, >5 kHz passband) are coaxially connected to the RF output end of the balanced detector, blocking low-frequency noise.

2) A wideband preamplifier (SR445A, Stanford Research Systems) provides further signal amplification. Depending on signal amplitude levels, up to 1000× amplification is attainable by cascading amplifier channels. Another DC block electric filter can be used after the pre-amplifier to further block the DC offset.

c. **Digitization and On-Board Averaging:** Data acquisition and logging are performed using a high-speed digitizer (GaGe RazorPlus, 16-bit, 500 MS/s) and software (GaGeScope Professional v.3.84.44). To avoid acquiring a large amount of data in the single record mode, we make full use of the multiple record mode and the on-board averaging function to visualize raw data, simplify data manipulation, and reduce the data size. Data is acquired and organized into a set number of record segments ($J$) upon receiving a rising-edge trigger event with a rate of a few kHz (see Figure S3). Typically, $J = 100$ multiple record segments are recorded, each containing $N$ data points. The data points include two time zero 'spikes' to encompass a complete scanning period. For example, $N = 6000$ may be acquired with a 10 MS/s sample rate. On-board averaging number $M = 1000$ is used to reduce the data size by three orders of magnitude in the digitizer memory. Tens of data files (I) are saved in the PC hard drive for further averaging and time-evolution analyses. The total number of signal averaging is $J \times M \times I$, and the total data points saved to the PC is $N \times J \times I$. Representative data averaging number, data file size, and data acquisition time are 1 million, 6 megabytes, and 1 minute, respectively (see section III for more details).

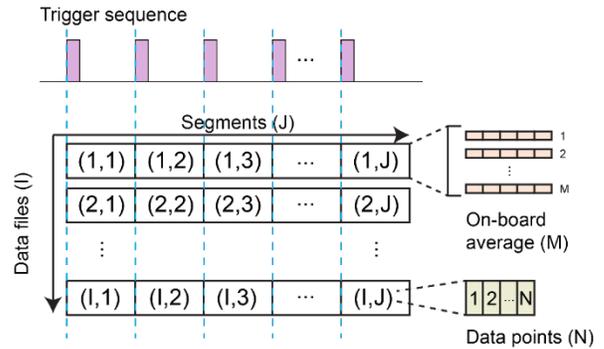

**Figure S3.** Trigger sequence and data structure.

d.     **Microscope:** A home-built microscope system based on a commercial microscope frame (RM21, Mad City Labs) is used for all measurements. The 1040 nm pump laser beam is steered by a silver mirror (M1) and a dielectric mirror (M2) to the bottom reflective objective (labelled as Obj.2, 74×/0.65, Beck Optronic Solutions), then focused to a diffraction-limited spot (~ 1 μm) in the sample plane. From the top, the 785 nm probe laser pulse is guided by mirrors M3-4 to the high numeric aperture objective (labelled as Obj.1, 100×/0.9, MPlanFL N, Olympus), which focuses the probe beam to ~1 μm spot. Diameters of both laser beams are also expanded to slightly overfill the objective lens. A half wave plate (HWP, Eksma Optics) and a quarter wave plate (QWP, Eksma Optics) are used to convert linear polarization to circular polarization for both laser beams. The polarization of the back-reflected 785 nm laser beam from the sample changes back to linear, now 90° shifted with respect to the incident beam. The back-scattering probe is separated by a polarizing beam splitter cube (labelled as PBS, CCM1-PBS252, Thorlabs) and directed to the detector. To observe and image nanoparticle samples, a blue LED light source (M490L4, Thorlabs) is collimated by a condenser lens (L2) and coupled to the top objective by a dichroic mirror (DM2) with a long pass cutoff at 506 nm (#67-080, Edmund Optics). A small portion of the back-scattering light passes through DM2, is reflected by another long pass dichroic mirror (DM1, FF700-Di01-25×36, Semrock), and imaged by an achromatic lens (L1, focal length=150 mm) onto a CMOS monochrome camera (BFS-U3-51S5M, Blackfly). Optionally, a white light LED (MWWHL4, Thorlabs) placed below the bottom objective may be used for imaging in transmissive mode. In addition, sample particles and the probe beam profile are imaged after a lens L3 with a focal length = 175 mm, and then relayed by a 4-f lens system consisting of lenses L4-5 with identical focal lengths (200 mm) to a 100 μm pinhole (PH). The spatially filtered light is imaged by an sCMOS camera (Zyla 5.5, Andor) via a flip mirror (FM) and an achromatic lens (L6, focal length = 100 mm).

**(II) Signal Isolation:** The raw signal in lab time for a single AuNP acquired by the digitizer is shown in Figure S4A. At 3 kHz offset, each scan completes in 0.333 ms. For a 4000 pt acquisition at 10 M/s, 0.400 ms of data are acquired, thereby revealing part of the next scan. This 20% overhead ensures that all the data is acquired and that the signal repeats exactly after each scan. The spike in the data corresponds nominally to time 'zero' where the two pulse trains overlap in time. Note, the frequency spectrum is unaffected by the exact time zero position. The signal exhibits complex time-domain oscillations which are a combination of two dominant effects: 1) the electronic conditioning which filters the signal using a high-pass electronic filter and a low-pass balanced detector, 2) the Raman signal arising from the acoustic vibrations of the nanoparticle including the axial, angular, and breathing modes. The first effect, primarily from the presence of the high-pass filter, causes a large distortion in the signal that must be corrected. This filter is necessary because the low-frequency components of the signal generated from the balanced detector overwhelm the small, coherent oscillations that encode the acoustic response. Without background subtraction, Fourier transforming the signal after the high-pass electronic filter buries the desired signal components. To remove the effects of the filter, we employed a Bayesian inference approach. For a 22 kHz high-pass filter, the expected frequency in the molecular frame is $v_{HP} = 22$ kHz$/\alpha$, where $\alpha = \Delta f/f_r = 3 \times 10^{-5}$ is the down-conversion factor. This gives $v_{HP} = 0.73$ GHz, which is still well below the expected acoustic frequencies of interest. In the Bayesian method, the entire signal is fit to a sum of exponentially decaying sinusoidal model functions, $g_i(t) = cos(\omega_i t + \phi_i)exp(-\Gamma_i t)$ in an iterative manner. Unlike many other nonlinear fitting methods, the amplitudes of the model functions are not directly fit, but rather treated as nuisance parameters using a variable projection approach. The nonlinear fitting algorithm uses a combination of local and global search methods, which are accelerated by explicitly calculating the Jacobian and Hessian of the model functions.

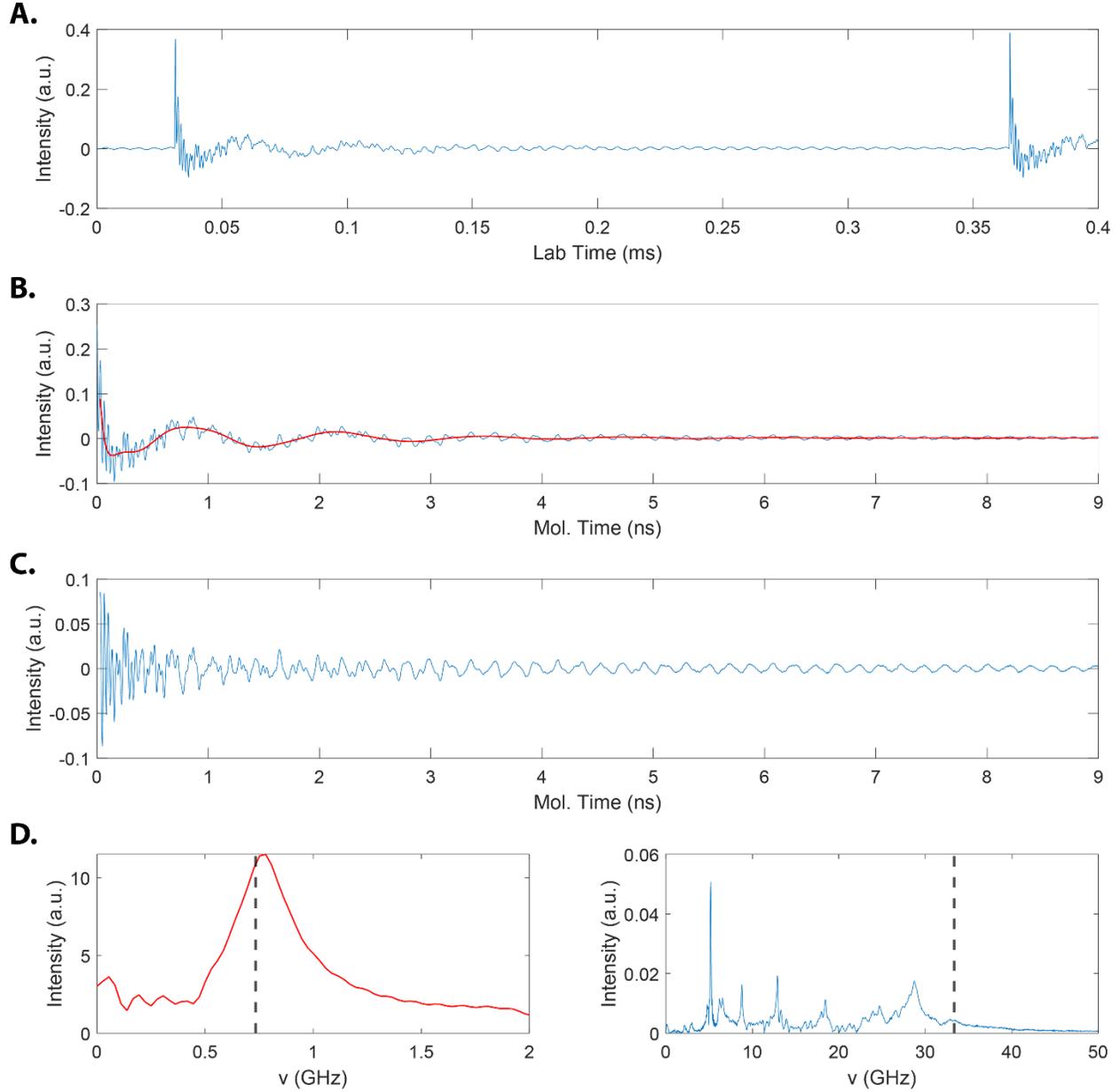

**Figure S4.** 100 nm AuNP background subtraction procedure. **A.** Original signal from digitizer. **B.** Signal in the molecular frame (blue) and low-frequency component of the fit (red). **C.** Residual of the signal in B. and its low-frequency fit. **D.** Left: FFT of the low-frequency fit (dashed line shows the expected center frequency of the high-pass filter at 22 kHz). Right: FFT of the residual in C. showing the acoustic modes. Dashed line shows the expected cut-off frequency from the balanced photodiode.

The algorithm finds the set of parameters, $\boldsymbol{\Phi} = [\omega, \phi, \Gamma]$, to minimize $Q(\boldsymbol{\Phi}) \equiv |S(\text{t}) - M(\boldsymbol{\Phi}, \text{t})B|^2$, where $S(t)$ is the signal, $M(\boldsymbol{\Phi}, t)$ contains the model functions as columns, and $B$ is a column vector of the amplitudes. Details of this approach can be found here (*29,37*). The iterative algorithm terminates when the standard deviation of the residual between the signal and the fit is at or below the estimated standard deviation of the noise. While this approach may be used to fit all signal components, here we only use it to remove the background signals below a prescribed cutoff frequency. For the signal shown in Figure S4B, we set $v_{cut} = 2.5$ GHz. The reconstructed fit for components with frequencies below the cutoff are shown in red, and the residual after subtracting the low-frequency signal is shown in Figure S4C. This signal, therefore, represents all the components with frequencies above the cutoff. As shown in Figure S4D, the background subtraction method corrects for the effects of the filter (red curve), retrieving the desired Raman signal (blue curve). To eliminate pulse-overlap effects and early-time dynamics from the spectrum, the analysis is started after 30 ps. This slightly degrades the spectral resolution of the breathing mode (by ~5-10%) and has a negligible effect on the axial modes (<1%). Note that the amplitude of the low-frequency components is ~200X that of the largest amplitude observed in the acoustic spectrum (near 5.2 GHz) and > 670X that of the breathing mode amplitude (near 28 GHz). There is also a low-pass filter that must be accounted for in the analysis, but its effect is much smaller. This filter is due to the limited bandwidth (~1 MHz) of the balanced detectors used here. In the molecular frame, this corresponds to a frequency of $v_{LP} = 1$ MHz/$\alpha$ = 33.33 GHz. Therefore, no signals greater than this frequency are detected even if the bandwidth of the pulses support up to ~2 THz. Using a higher bandwidth detector or a lower offset frequency would allow a wider range of acoustic frequencies to be detected. For instance, for a 10 MHz bandwidth detector and a 1 kHz offset, signals as high as 1 THz could be measured.

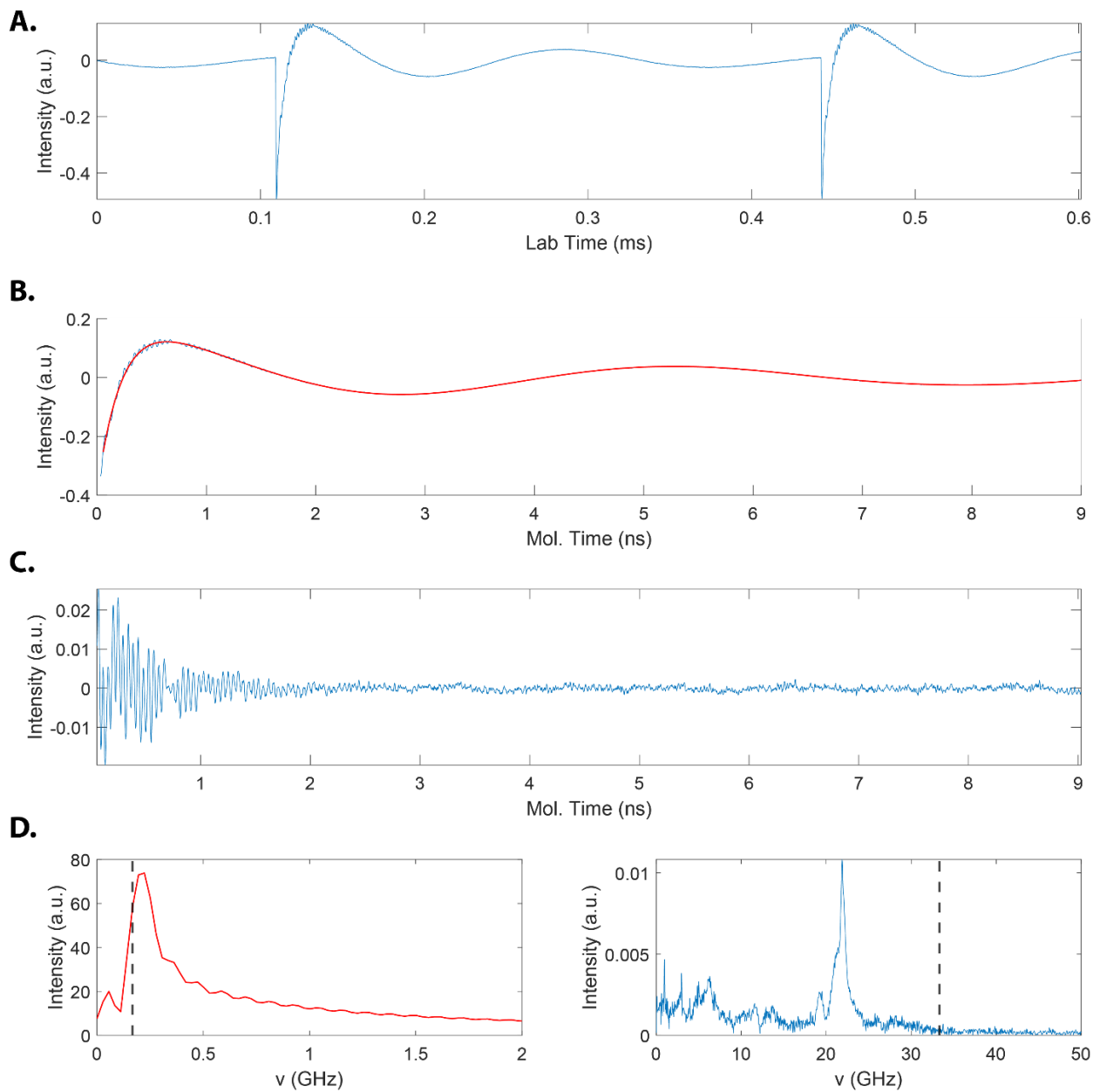

**Figure S5.** LentiGFP background subtraction procedure. **A.** Original signal from digitizer. **B.** Signal in the molecular frame (blue) and low-frequency component of the fit (red). **C.** Residual of the signal in B. and its low-frequency fit. **D.** Left: FFT of the low-frequency fit (dashed line shows the expected center frequency of the high-pass filter at 5 kHz). Right: FFT of the residual in C. showing the acoustic modes. Dashed line shows the expected cut-off frequency from the balanced photodiode.

The same analysis applied to one of the trajectories for a single Lentivirus is shown in Figure S5. Note that here we used a 5 kHz high-pass electronic filter ($v_{HP} = 0.17$ GHz) but all other parameters remained the same.

**(III) Estimating the Acoustic Frequencies (Lamb's Model)** The set of vibrational frequencies of a spherical particle can be estimated using an elastic continuum theory under stress free boundary conditions (*36*). The equation of motion according to Lamb for a three-dimensional elastic body is given by $\rho \frac{\partial^2 \vec{D}}{\partial t^2} = (\lambda + \mu)\nabla(\nabla \cdot \vec{D}) + \mu \nabla^2 \vec{D}$, where $\vec{D}$ is a lattice displacement vector, μ is the Lame's constant and ρ is the mass density. Using knowledge of the bulk longitudinal and transverse sound velocities, $V_l$ and $V_t$, respectively, it is possible to calculate the energy eigenvalues for an isotropic spherical particle using stress free conditions at the surface. These eigenvalues are described by an orbital angular momentum quantum number, $l$, and a harmonic, $n \geq 1$. Only the even $l$ modes are Raman active for the spherical modes with $l = 0$ correspond to purely radial motion with spherical symmetry, while $l = 2$ correspond to quadrupolar motion. The torsional modes are not Raman active for a free, spherical particle, but are listed below as a break in the particle symmetry may impart a non-zero transition probability to these modes. For a free, spherical particle, $l = 0, 1, n = 1$ are surface modes, while $n \geq 2$ are inner modes. Note, that in some references the harmonics are enumerated starting at $n = 0$, while here we use the convention starting at $n = 1$. When the particle deviates from spherical symmetry, the degeneracy lifts and each mode splits into $l + 1$ modes.

The longitudinal and transverse velocities of a virus particle are usually assumed to be close to that of lysozyme protein crystal: $V_l = 1817$ m/s, $V_t = 915$ m/s. For an $D = 72.5\ nm$, spherical virus, the normal mode frequencies in GHz units are given in Table S1.

**Table S1.** Normal modes frequencies for a 72 nm spherical virus using the sound velocity of a lysozyme protein crystal. Orange shaded boxes are modes that are outside the detection bandwidth. Purple shaded boxes are for modes that are not Raman active for a free, spherical particle. Bold values are modes that are close to the measured values (within 10%).

| Spheroidal | | | Torsional | | |
|---|---|---|---|---|---|
| $l$ | $n$ | $\nu$ | $l$ | $n$ | $\nu$ |
| 0 | 1 | **21.81** | 0 | 1 | - |
| 0 | 2 | 48.75 | 0 | 2 | - |
| 1 | 1 | **14.41** | 1 | 1 | 23.13 |
| 1 | 2 | 29.03 | 1 | 2 | 36.52 |
| 2 | 1 | 10.64 | 2 | 1 | 10.03 |
| 2 | 2 | **20.43** | 2 | 2 | 28.66 |

Recent inelastic Brillouin light scattering experiments estimate that the sound velocity in a virus may be up to ~15% lower than in the lysozyme crystal due to the higher density of the protein capsid and nucleotide core. This would translate to a 15% correction in the virus diameter, giving $D = 83$ nm. For a 105 nm AuNP ($V_l = 3240$ m/s, $V_t = 1200$ m/s) the normal mode frequencies are given below.

**Table S2.** Normal mode frequencies for 105 nm AuNP. Orange shaded boxes are modes that are outside the detection bandwidth. Purple shaded boxes are for modes that are not Raman active for a free, spherical particle. Bold values are modes that are close to the measured values (within 10%).

| Spheroidal | | | Torsional | | |
|---|---|---|---|---|---|
| $l$ | $n$ | $\nu$ | $l$ | $n$ | $\nu$ |

| | | | | | | |
|---|---|---|---|---|---|---|
| 0 | 1 | **28.92** | 0 | 1 | - | |
| | 2 | 60.83 | | 2 | - | |
| 1 | 1 | **13.6** | 1 | 1 | 20.95 | |
| | 2 | 26.84 | | 2 | 33.07 | |
| 2 | 1 | 9.62 | 2 | 1 | 9.09 | |
| | 2 | **18.51** | | 2 | 25.95 | |

We note that the measured shearing frequencies (angular mode) shown in Figure 2A of the main manuscript are close to, but not exactly, those predicted above: 14.55 (weak) and 19.31 GHz. The latter may be the $l = 2, n = 2$ spherical mode, while the former is closest to the $l = 1, n = 1$ spherical mode which is not Raman active for a free particle but may be weakly allowed when the spherical symmetry is broken. However, as these modes are coupled to the axial modes, the Lamb model is not directly applicable, and a more advanced analysis is needed. Improved assignments may be made by calculating the Raman intensity which should be proportional to the mean square displacement ($\mu^2$), but that was beyond the scope of the current study(35).

**(IV) Sample Preparation:**

**Gold Nanoparticles:** Glass slides were ultrasonically cleaned in 1 M sodium hydroxide (NaOH) solution, Milli-Q water, and ethanol for 30 min, respectively. The glass slides were immersed in piranha solution ($H_2SO_4:H_2O_2$, 5:1) for 10 min to decorate the glass substrate with hydroxyl groups ($OH^-$). For the tethered AuNPs, the glass slides were incubated in 5% (V/V) (3-aminopropyl) triethoxysilane (APTES) in ethanol for 3h. After 3h the cover glass was dried under $N_2$ stream. The slides were then subjected to thermal annealing in a vacuum oven at 110 °C for 2 h to obtain APTES-silanized slides with amine groups, which were subsequently immersed into the AuNPs solution (Sigma-Aldrich, #742031, 100 nm, stabilized suspension in citrate buffer, <0.2 polydispersity index) for 6 h adsorption. The amino groups on the APTES molecules are used to immobilize gold particles onto the substrate due to the specific affinity of the amino group to the colloidal gold nanoparticles (*38*). After the Au NP adsorption, the glass slides were washed and dried under $N_2$ stream. For untethered AuNPs, the sample was simply drop cast onto the glass slide. The citrate-capped AuNPs readily absorb onto the (OH)-functionalized glass surface due to a combination of van der Waals forces (via the citrate ions) and hydrogen bonding interactions (via carboxylate groups on the citrate).

**Lentiviral Pseudovirions:** Lentiviral pseudovirions were produced in 293FT cells by transfecting with packaging constructs pCMV-VSV-G and pCMV-Delta 8.2 (gifts from Jerome Schaack). Cell culture supernatant containing virions was collected 72 hrs post-transfection and filtered using a 0.45µm filter to remove cell debris. The virions in the supernatant were concentrated by ultracentrifugation at 20,000 rpm for 2 hrs. The pelleted virions were inactivated and fixed using 4% paraformaldehyde.

**(V) Other 100 nm AuNP Trajectories:** Here we show two additional AuNP trajectories that show similar acoustic spectra to those in the manuscript. In Figure S7, the trajectory for an untethered, single AuNP is shown, displaying spectral shifts with similar trends to those in Figure 2B of the manuscript. The acoustic spectra, in general, exhibited more variability among the untethered AuNPs compared to the tethered samples because of local environment effects. The extent of spectral shifting is highly dependent on the particle environment. In Figure S8, the trajectory for a tethered, single 100nm AuNP is shown, exhibiting similar spectral features to those shown in Figure 2A (manuscript). While the breathing mode for most AuNP measured was around 25 - 28 GHz, the line shapes differed significantly. This is due to different factors such as the presence of multiple crystalline domains, deviations from spherical symmetry, and matrix effects.

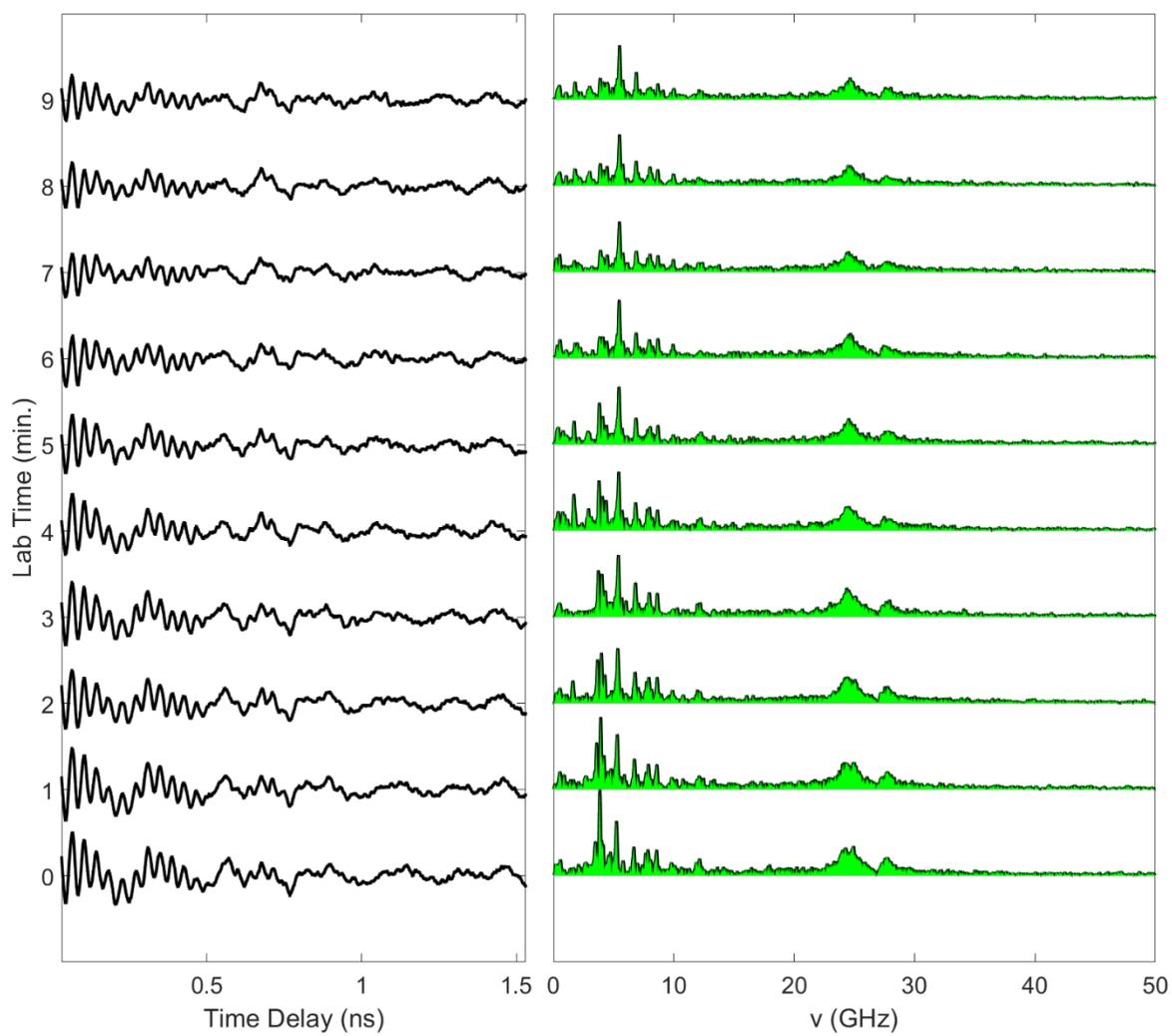

**Figure S6.** Untethered, single AuNP trajectory.

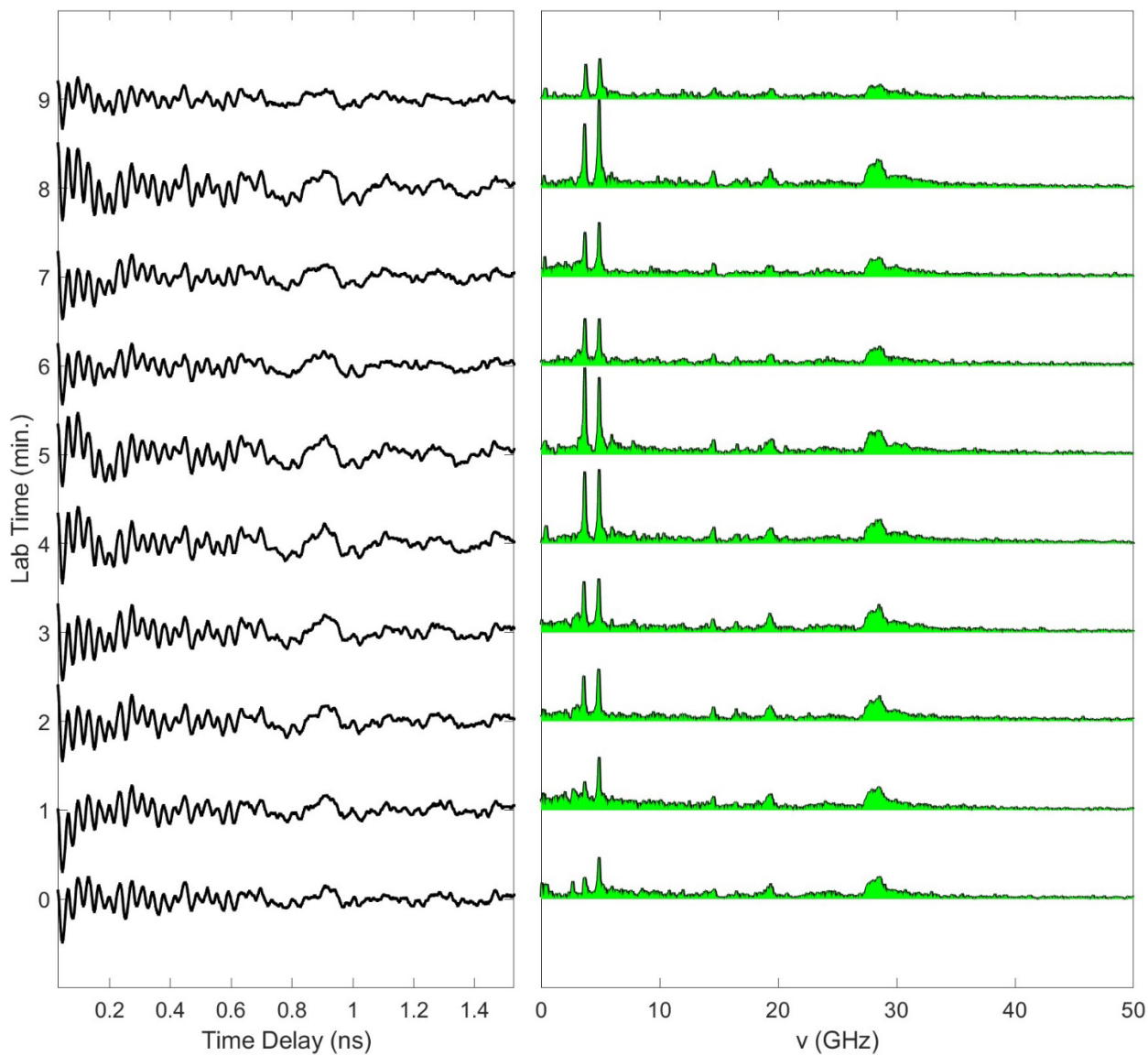

**Figure S7.** Tethered, single 100nm AuNP trajectory.

**(VI) Signal-to-Noise Ratio and Time Resolution:** The trajectories shown in the main manuscript are displayed in increments of 1 minute, but the time resolution of the measurement may be higher depending on the signal-to-noise (SNR) ratio. Since the digitizer performs 1000 on-board averages, each measured trace corresponds to 0.33 seconds of acquisition time. Only about 80% of the signal is used in the analysis, so that each trace shown corresponds to ~0.25 seconds of acquisition. For AuNP, the signal is sufficiently strong that one trace (×1) generates high SNR > 10 (see Figure S6). The on-board averaging may be reduced further in this case to enable millisecond acquisition times. For the viruses, however, the SNR is significantly weaker. Shown in Figure S6 is the same analysis applied to a single virus particle. Here, at least 10 trace averages (10,000 total averages) are needed for an SNR >10 for the breathing mode, while the axial mode scarcely rises above the noise floor. At this level of averaging, the time resolution is ~2.5 seconds. If only the breathing mode is required, the time resolution may be as low as 0.25 seconds. Note, that in some figures, we display vertical dashed lines which indicate the position of noise peaks arising from the timing electronics, amplifiers, and detector electronics. These features are usually well below the signal level, but for low SNR traces, they manifest as sharp spectral features (0.1 GHz line width) because they do not decay in time.

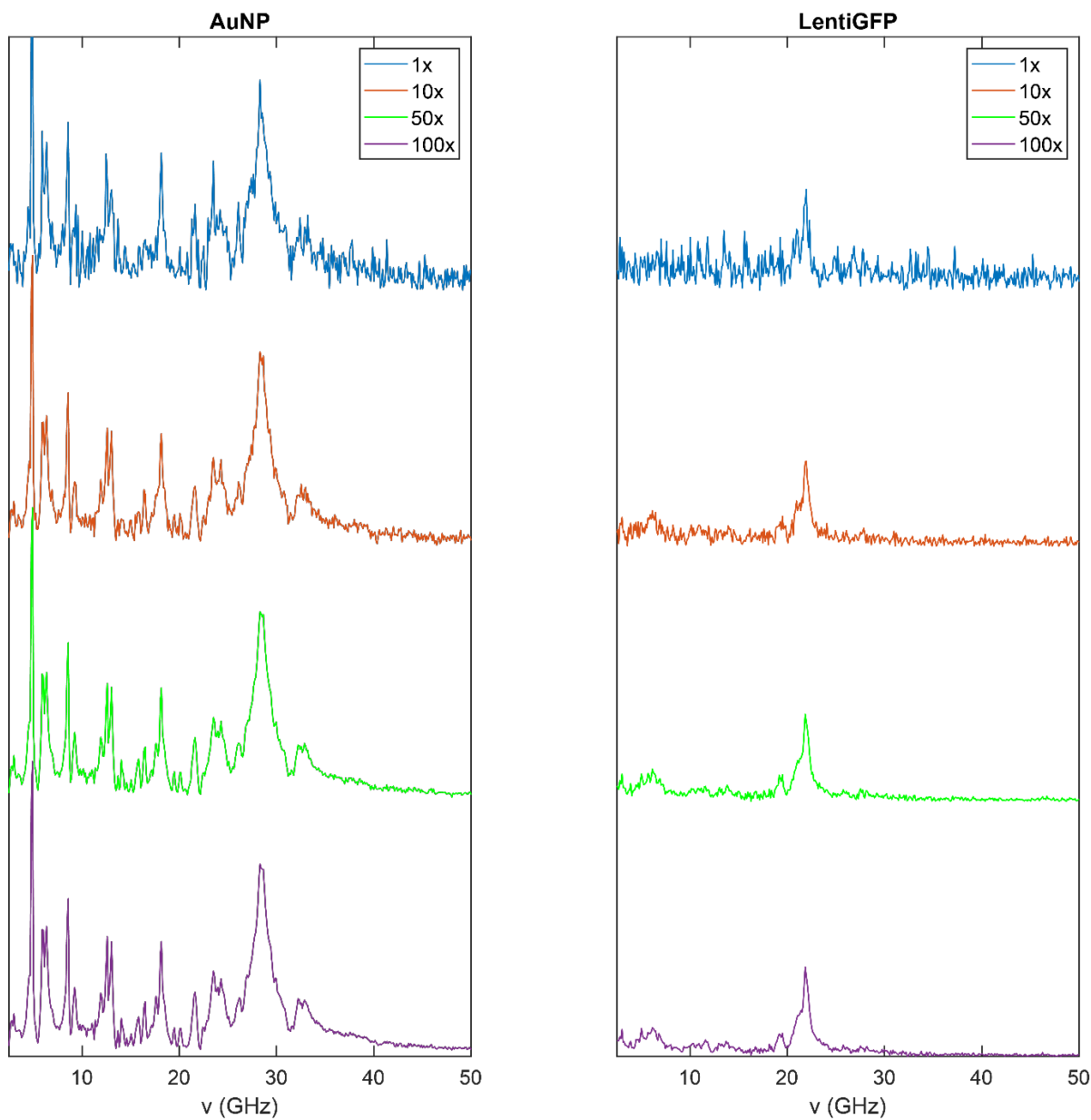

**Figure S8.** Effect of signal averaging for single AuNP and single Lenti virion. Note, each trace corresponds to 1000× on-board averaging. Further averaging is shown in the legend (×1, ×10, ×50, ×100).

**(VII) Other Lentivirus Trajectories:** Shown in Figure S10 is the trajectory of Virus #3. Here, the axial mode is largely absent, while the breathing mode exhibited small spectral shifts and line shape changes as a function of experiment time. This may be due to the virus undergoing slight deviations from spherical symmetry (see section IX). A fourth trajectory (Virus #4) is shown in Figure S11. In this case a weak axial mode is observed, while the spectra remain largely unchanged throughout the trajectory. A fifth trajectory (Virus #5) is shown in Figure S12, whereby broad axial and shearing (angular) modes are observed. According to the AFM images, Virus #1 (manuscript) and Virus #3 are free particles on the substrate, while Viruses #2, #4, and #5 are buried to varying degrees. This is consistent with the observation of stronger axial modes which are due to virus/matrix interaction.

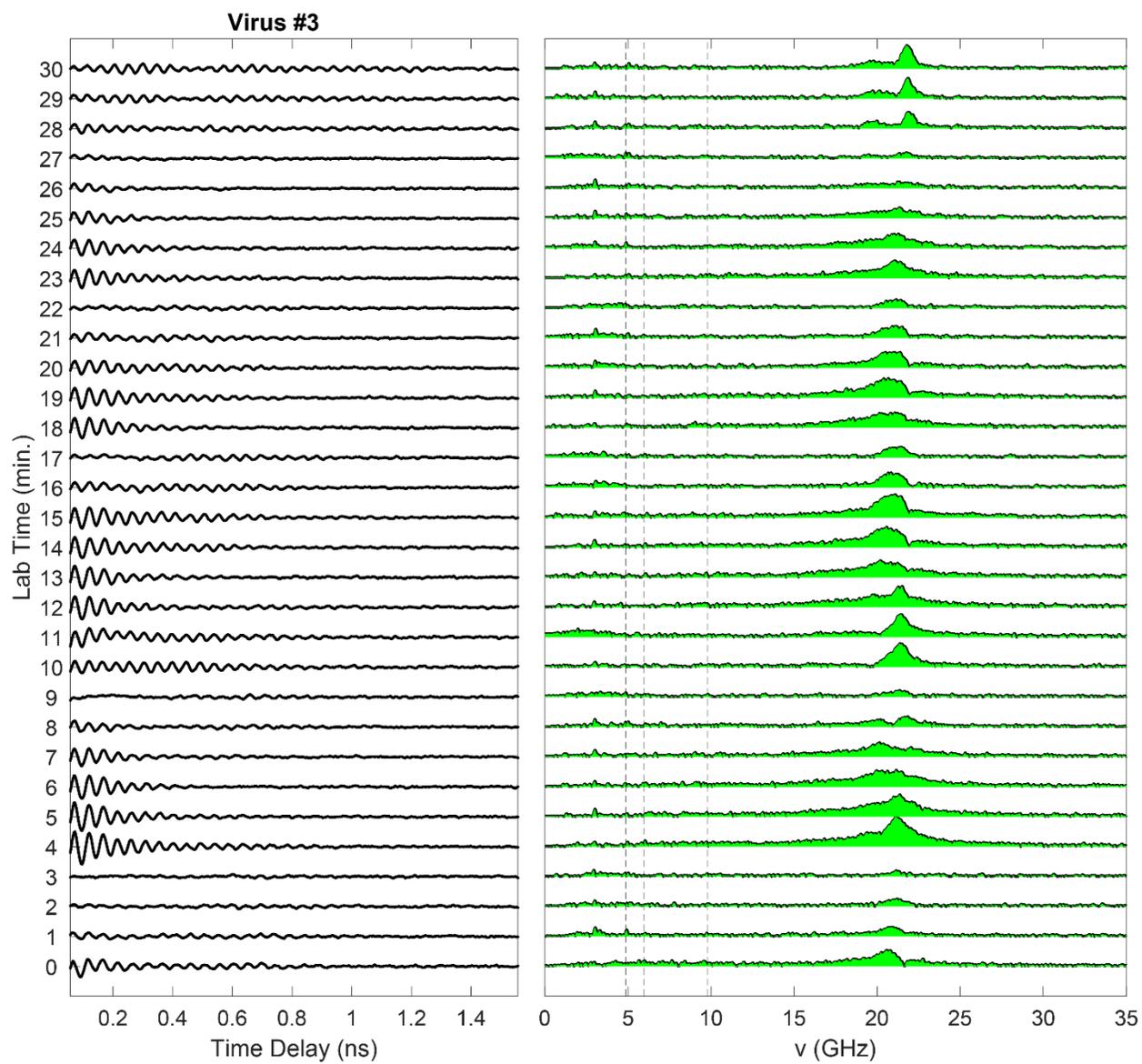

**Figure S9.** Trajectory of Virus sample #3 showing partially reversible dynamics. See also a discussion in Section (IX)

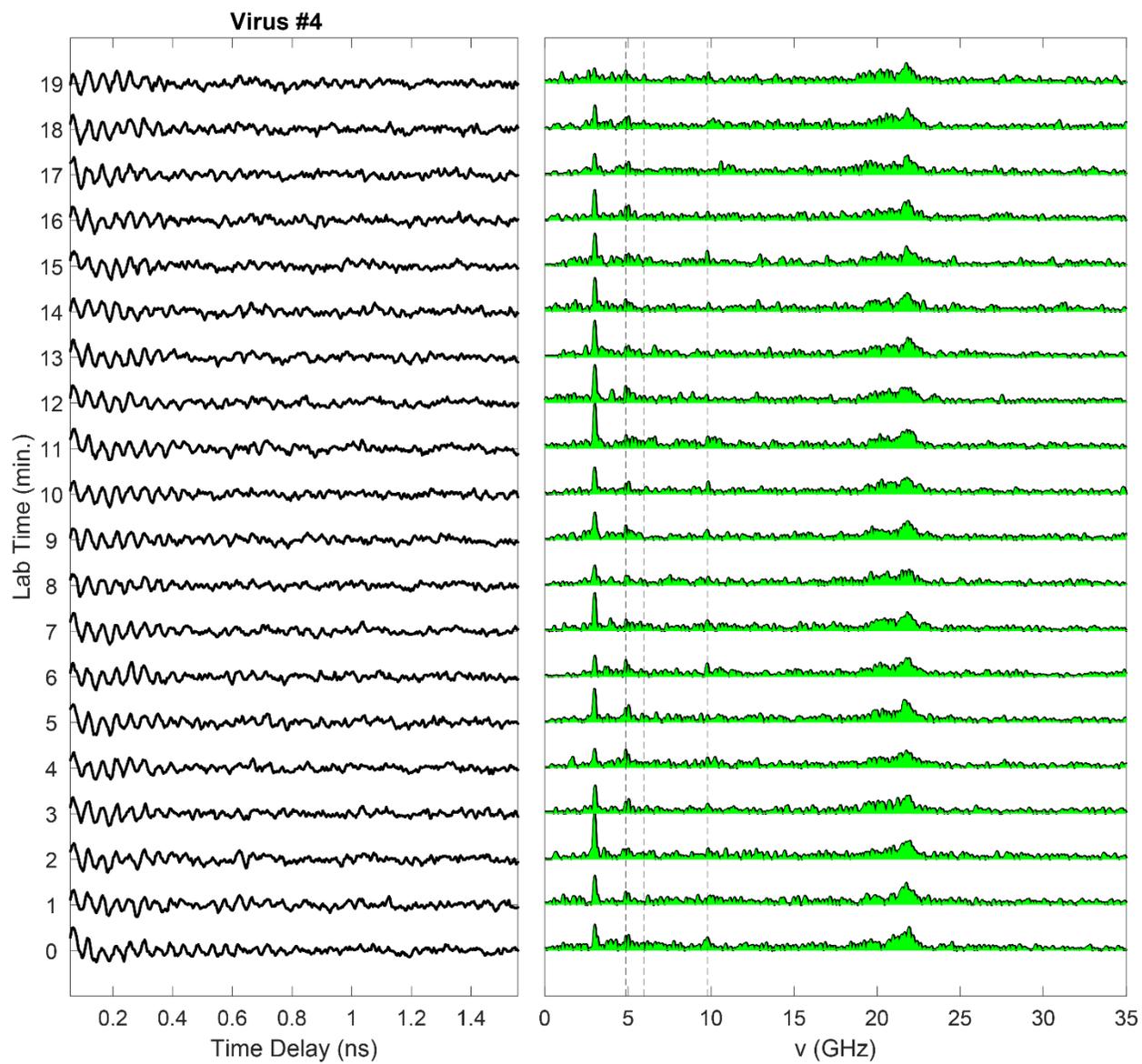

**Figure S10.** Trajectory of Virus sample #4 showing stability of the particle over 20 minutes.

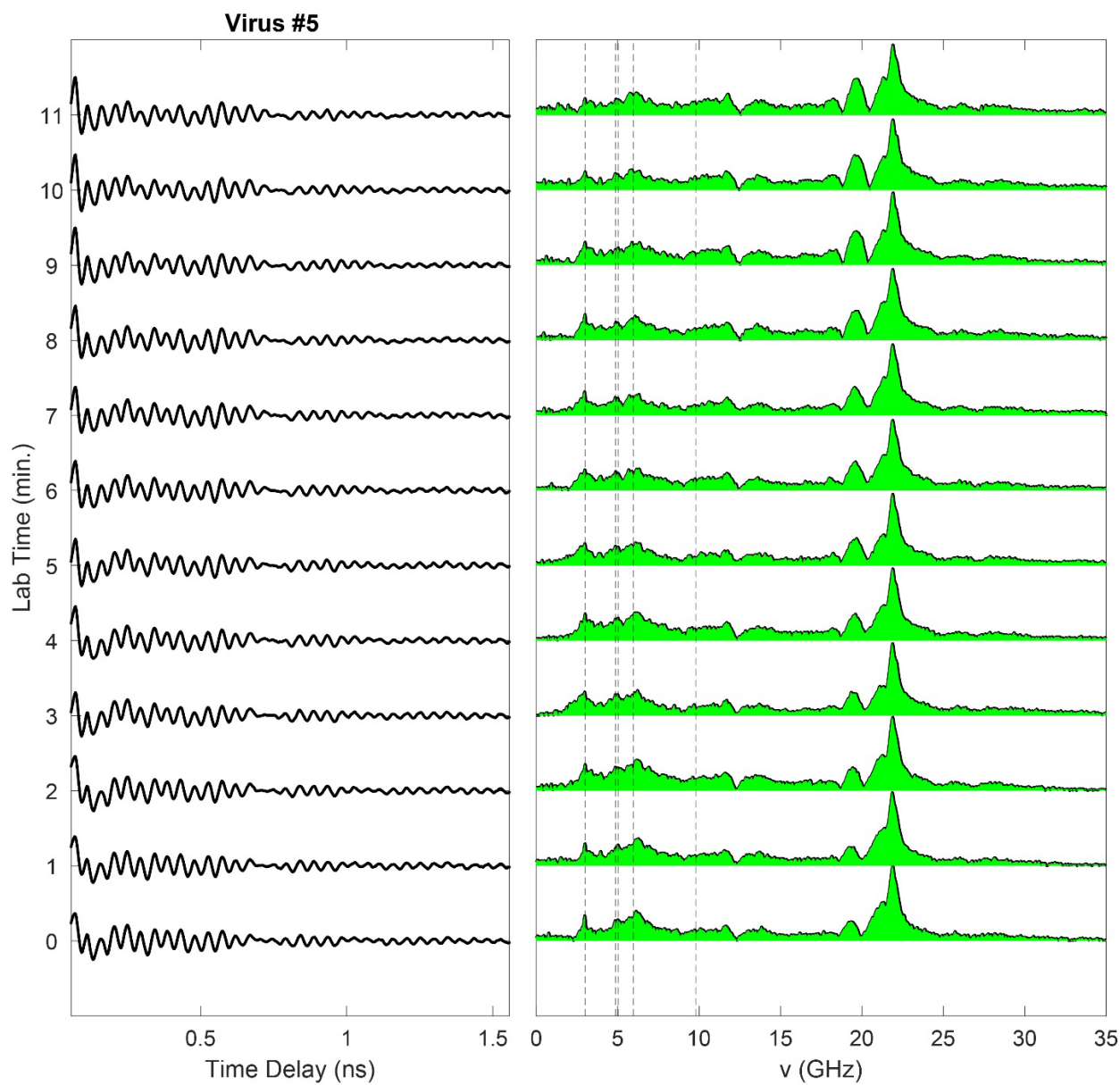

**Figure S11.** Trajectory of Virus sample #5 showing prominent axial and angular modes.

**(VIII) Correlative Atomic Force Microscopy (AFM):** For AFM experiments, gridded glass coverslips (Grid-50, ibidi) were first plasma cleaned for ~ 10 minutes, then washed by ethanol and distilled water. After coverslips dried in air, a drop of ~ 5 µL virus solution (4% paraformaldehyde fixation) was introduced to the gridded surface of these coverslips. Five minutes were allowed for the virus adsorption onto the surface at room temperature, followed by washing with ~ 20 µL distilled water to stop formaldehyde crystallization. For mica substrates, freshly cleaved mica sheets (Grade V-1, Electron Microscopy Sciences) were used, and all other subsequent processes are similar to those used for the glass slides.

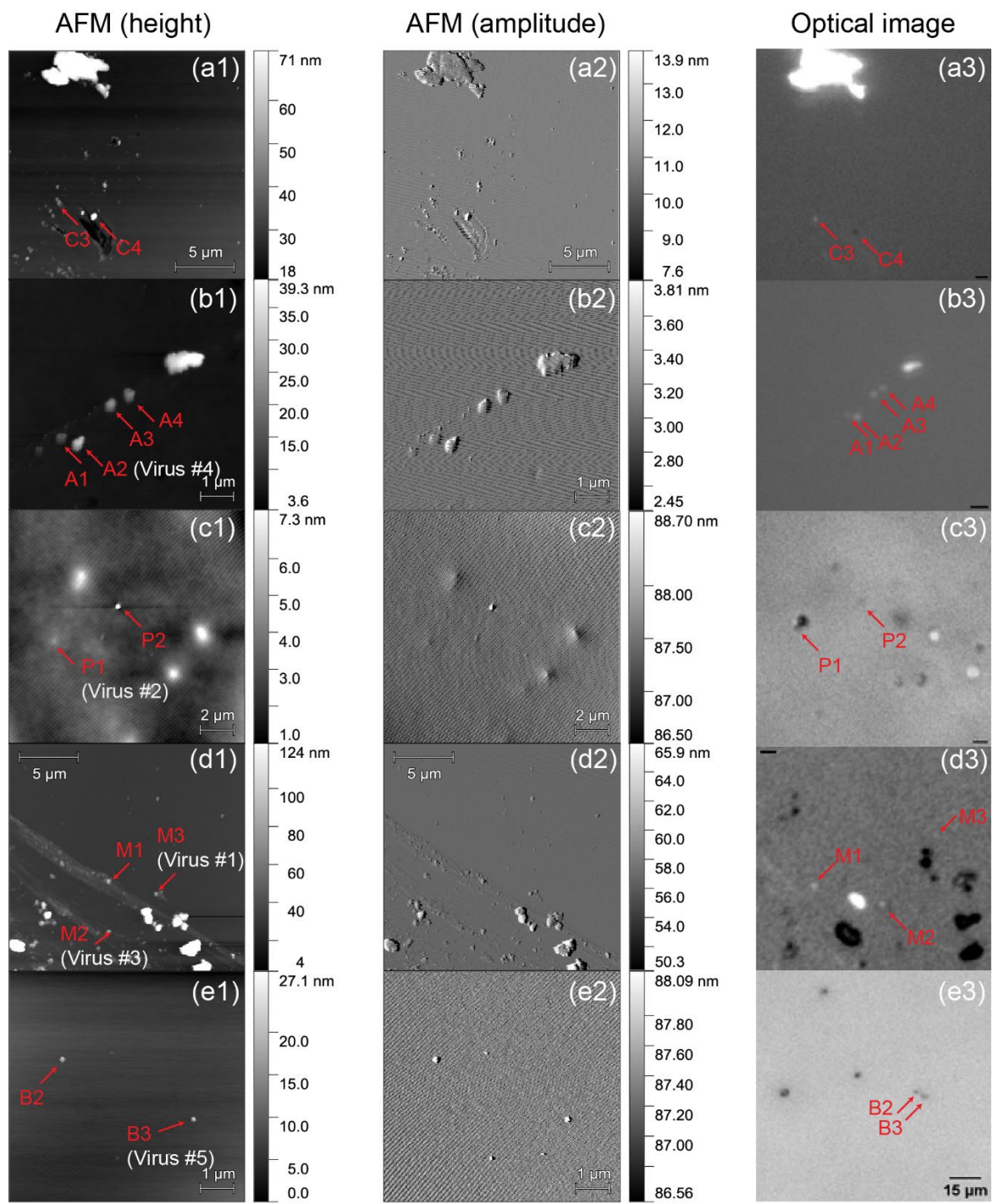

**Figure S12.** AFM images and correlative optical images. The left column shows common height profile images of these virus particles. These images are best to show the topography and height information. The middle column shows the amplitude images (due to AFM tip deflection). These images can show better shapes and surface information. The right column shows optical images. Note, the system response causes lateral broadening and non-spherical shapes for some cases (e.g. in (b2)). Due to very high (~ 1μm) markers, some virus particles may slightly move out of position (see M1 in (d1) and (d3)). Scale bar is 1 μm if not indicated.

AFM images were acquired using a high-speed, multimode scanning probe microscope (Cypher S, Oxford Instruments Asylum Research). AFM probe tips were 160 μm long, had a spring constant of 26 N/m, and an experimental resonant frequency of 265.9 kHz (AC160TS-R3, Oxford Instruments). AFM scanning was operated in tapping mode to reduce sample damage, which is achieved by the intermittent contact of the probe tip with the sample surface. Specifically, we scanned a 20 μm × 20 μm area of interest at a 0.5 Hz scan rate to find previously registered virus particles using optical microscopy; then we zoomed into the area containing virus particles and ran fine scans at a 0.1 Hz scan rate across a 5 μm × 5 μm or 1 μm × 1 μm region. Offline data analyses was done in the commercial software Gwyddion. Monodispersed 100 nm gold nanospheres (Nanopartz) were used to compare and correct the lateral dimension.

To ensure that single particles were measured, rather than clusters, we correlated AFM images with optical images of the same virus particles (see Figure S9). Etched labels on cover slips, patterns formed by salt, paraformaldehyde crystals, or sample particles were used as markers in the AFM and optical images to locate the sample particles in the respective microscopes. Note that due to the buffer and impurities in the sample, the virus particles are oftentimes embedded in a matrix, rather than being freely situated directly on the glass substrate. Therefore, the heights recorded in the AFM do not necessarily represent the height of the free particle.

**(VIV) Underdamped Harmonic Oscillator:** A damped harmonic oscillator may be described according to the second order differential equation: $m\ddot{u}(t) + c\dot{u}(t) + kx = 0$, where $m$ is the mass, $c$ is the damping coefficient, $k$ is the spring constant, and $u(t)$ is the time-dependent displacement of the mass. Defining the damping ratio as $\eta = c/2\sqrt{mk}$, the solutions for the underdamped case ($\eta < 1$) are given by $u(t) = e^{-(c/2m)t}(A\cos(\omega_d t) + B\sin(\omega_d t))$, where $A$ and $B$ are determined by the initial conditions and $\omega_d = \omega_0\sqrt{(1-\eta^2)}$, with $\omega_0 = \sqrt{(k/m)}$ corresponding to the solution for the undamped case. In the experiment $\omega_d$ and $\Gamma = \eta\omega_0$ are measured for each resolvable mode, which may then be used to calculate $\omega_0$ and $\eta$ for each mode. For the AuNP in Figure 2B, the dephasing rate of the breathing mode decreases from 1.02 GHz to 0.62 GHz during the 9-minute trajectory (39% drop in the damping coefficient). Therefore, the damping ratio for the breathing mode is $0.022 < \eta_{br} < 0.036$, and for the axial mode $0.020 < \eta_{ax} < 0.042$, well within the limits for the underdamped model. The small damping ratio implies that $\omega_d \approx \omega_0$. The change in stiffness during the trajectory may be estimated by $\Delta k = m[\omega_0^2(T_f) - \omega_0^2(T_i)]$, where $T_{i(f)}$ is the initial(final) time point in the trajectory. Alternatively, if $k$ is constant, then the change in relative effective mass may be estimated by $\Delta m/m = [\omega_0^2(T_f) - \omega_0^2(T_i)]/\omega_0^2(T_f)$. For virus particle #2, (Figure 2B in main manuscript), the dephasing rate of the breathing mode decreases from 1.43 GHz to 0.90 GHz during the measured trajectory. Therefore, the damping ratio for the breathing mode is $0.030 < \eta_{br} < 0.064$, and for the axial mode $0.032 < \eta_{ax} < 0.075$, well within the limits for the underdamped model. Again, the small damping ratio implies that $\omega_d \approx \omega_0$.

**(X) Partially Reversible Dynamics under Weak Virus/Environment Interaction:** A single virus may present partially reversible dynamics in the limit of weak virus/environment interaction. As shown in Figure S13 , spectra were recorded for 15 minutes, then the laser was blocked for 10 minutes (minutes 16 – 25) to avoid excess heat accumulation, and finally another 10 time traces were acquired for minutes 26 – 35. The weak intensities of the axial modes in the 0 – 10 GHz range indicated that the virus particle had a weak interaction with its environment, while the breathing mode persisted for all the spectra and its line shape evolved with time. We observed that the shape and amplitude change of the last 7 spectra (blue) largely resemble those of the first 7 spectra (red). This observation could be ascribed to the reversible dynamics of a virus particle: the vibration of a particle at a late time restored to that at an earlier time. For example, both the time response and spectrum at minute 29 (blue) almost reproduces those at minute 0 (red), see Figure S13(c-d). Similar spectral shapes are also observed in minutes 4 -6 and minutes 33 – 35. Lentivirus is known to have a pleomorphic shape (i.e. changes shape), but it is predominantly elliptical or spherical (*35*). As viruses can change shape due to thermal fluctuations, it is possible that some local heating effect of the matrix due to the prolonged laser irradiation caused a slight shape change. We also noticed that an appreciable discrepancy in some pairs of spectra, e.g. spectra at minute 3 and minute 32, causing an incomplete reversibility. In some traces, the spectra are clearly split between the main breathing mode (~ 22 GHz) and a side mode (~ 18 GHz), which is likely the result of an asymmetric deformation of the virus.

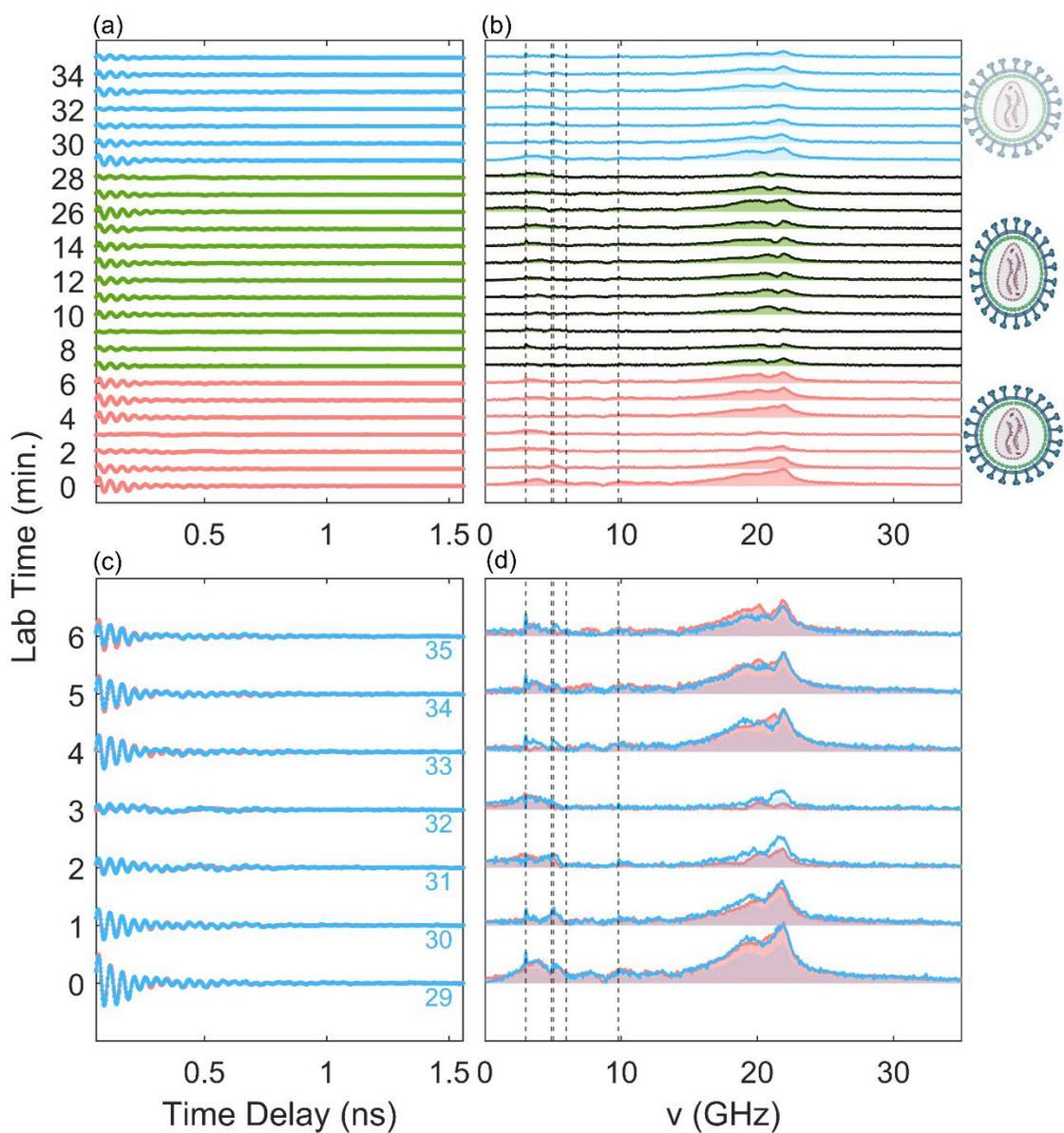

**Figure S13.** Partially reversible dynamics of a lentivirus in the limit of weak virus/environment interaction. (a) 26 time traces recorded in minutes 0 -15 and 26 -35. In minutes 16 – 25, the laser is blocked. (b) Extracted spectra associated with data in (a). (c – d) Enlarged view of data in minutes 0 – 6 (red) and 29 – 35 (blue), illustrating the partially reversible dynamics. In (c-d), for a better comparison, a scale factor of 1.44 is used for the spectra in minutes 29 – 35 (blue).

**(XI) Gold Nanourchin - Spiky NP:** To provide further evidence that the spike glycoprotein plays contribute to the axial mode splitting, we measured a series of time traces and spectra from a single ~ 100 nm gold nanourchinm, which are multibranched nanoparticles with a spiky uneven surface. Although the nanourchin is composed of a metal which has very different properties to the molecular components of a virus, the biomimetic spiky surface of the nanourchin shares some mechanical similarities that we wished to highlight. The spectral trajectory of the nanourchin clearly illustrates that there are multiple spectral peaks in the axial mode in the range of 2 – 10 GHz, a feature similar to those in a lentiGFP virus (see Virus#2, Figure 3; Virus#5, Figure S16). The breathing modes at ~ 30 GHz are weak, presumably from the low symmetry of the nanourchin shape, but the axial modes are strong and persist in all the spectra. The broad linewidth for axial mode peaks in the 2 – 5 GHz region suggests associated vibrations experience more damping (i.e. due to local environments) than those in the 5.5 – 10 GHz region.

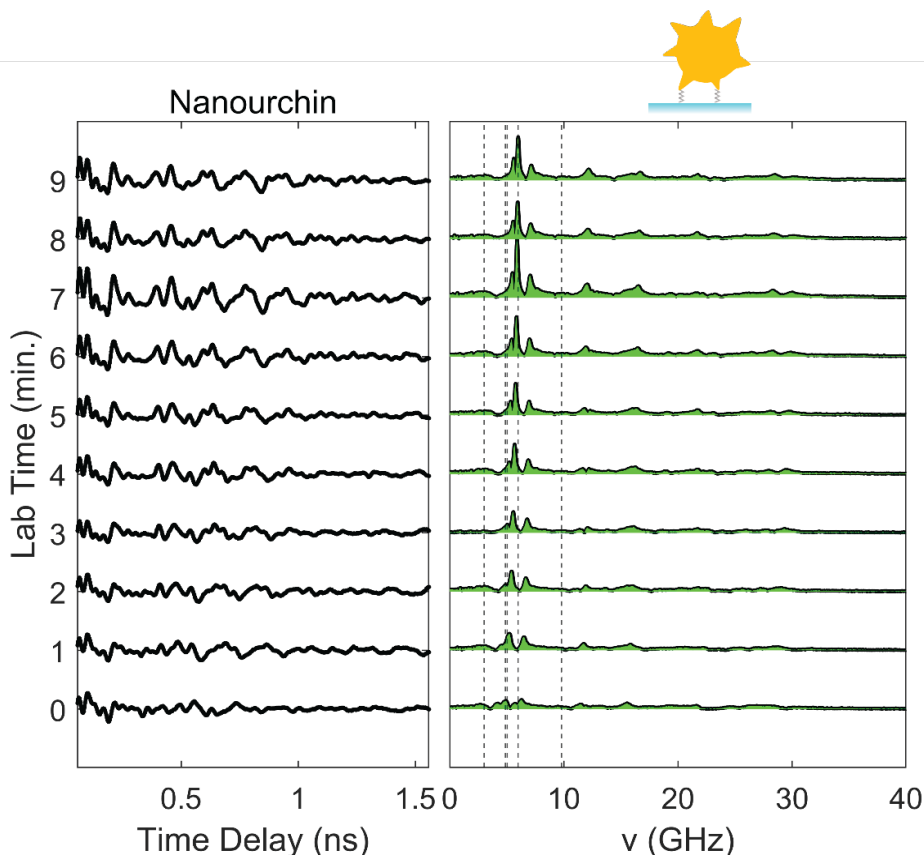

**Figure S14.** Trajectory of a ~ 100 nm gold nanourchin. Inset on the top right side: illustration of a gold nanourchin,

**(XII) Lentivirus and AuNP on the Mica Substrate:**

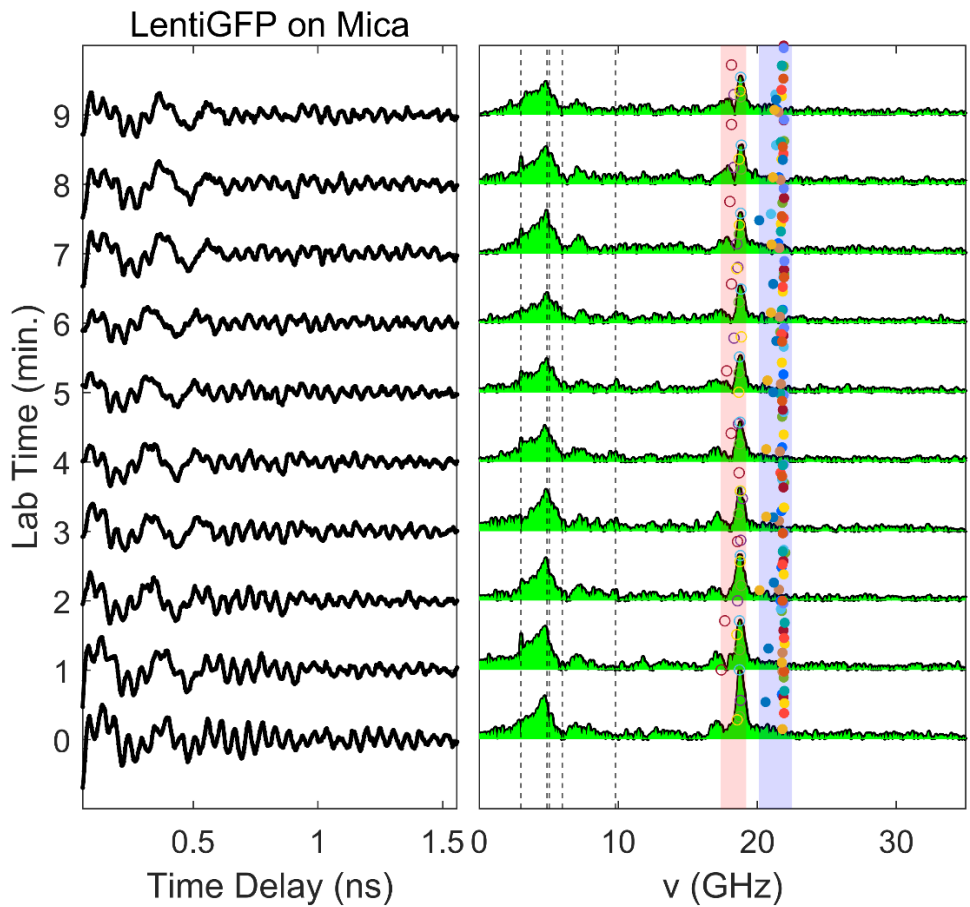

**Figure S15.** Time trace and spectral trajectory of a lentiGFP virus on a mica substrate. Trajectory is recorded for 10 minutes. Open circles in the red shade area show breathing mode peaks of single lentiGFP virus particles, and solid circles in the blue shade area are breathing mode peaks of the lentiGFP virus but deposited on fused silica cover glass. 50 (120) breathing mode peaks of 5 (12) viruses on mica (cover glass) are included. A redshift of ~3 GHz for the breathing mode peak is observed.

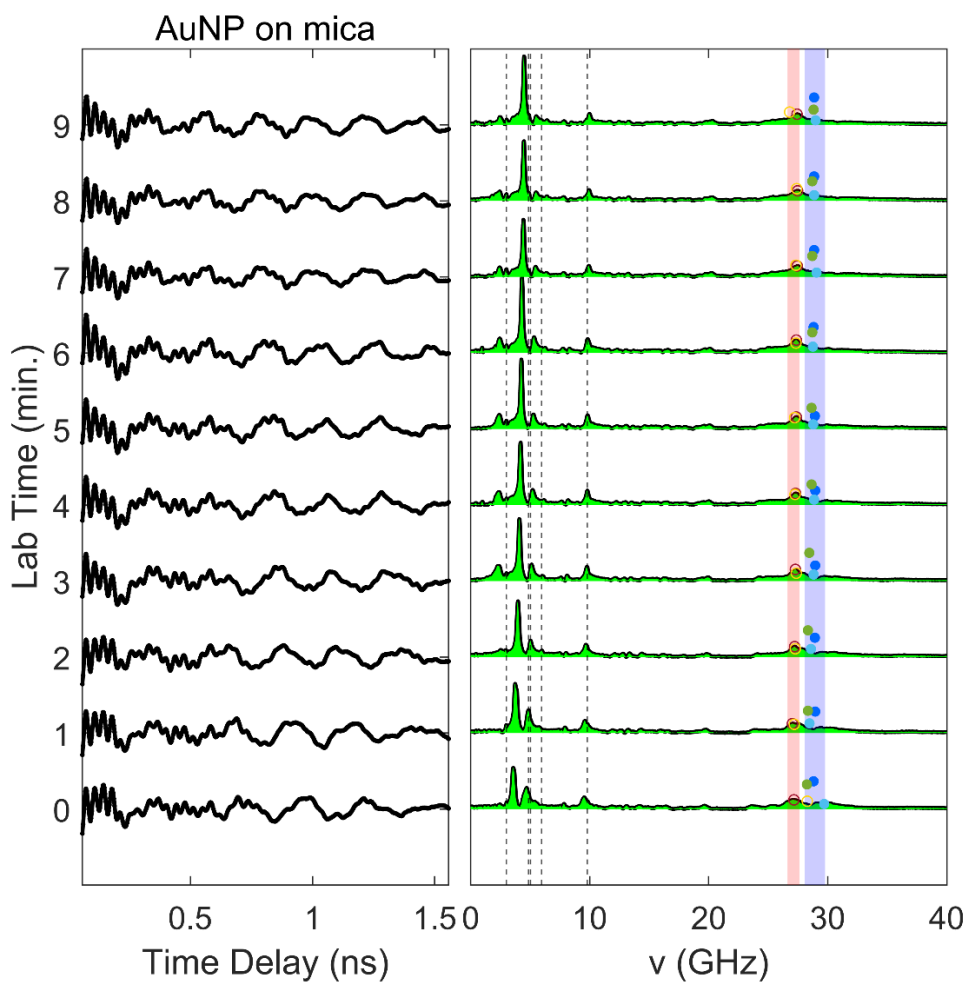

**Figure S16.** Trajectory of a 100nm gold nanoparticle on the mica substrate. Trajectory is recorded for 10 minutes. Open circles in the red shade area are the breathing mode peaks for 20 spectra from 2 particles. For comparison, solid circles in the blue shade area are the breathing mode peaks for 30 spectra from 3 particles, but deposited on a fused silica cover slip. Despite size and shape heterogeneities, there is a ~ 1.5 GHz red shift for the breathing mode with the mica substrate. There is also a discernable blue shift for the observed modes.

**(XIII) Statistical Analysis:** To perform statistical analysis on the virus data sets, we performed Bayes analysis on each of the $N = 212$ time-domain traces. The traces were analyzed from 30 ps to 9 ns to avoid including early-time dynamics. The algorithm used a maximum of 25 iterations where the parameters were bound within the range of $v \in [0,40]$ GHz, $\Gamma \in [0,10]$ GHz, and $\phi \in [0,2\pi]$. The initial point was set to $\mathbf{\Phi} = [0, 0, 0]$. The global search algorithm used with a simulated annealing (Mathworks, Matlab 2023, Global Optimization Toolbox), while the local search algorithm was a constrained nonlinear optimization with a trust region reflective algorithm (Mathworks, Matlab 2023, Optimization Toolbox). For the latter, analytical forms of the Jacobian and Hessian of the objective function were supplied. The Bayes algorithm extracts a maximum of 25 frequencies, decay rates, and phase terms for each trajectory. This value was chosen because the standard deviation of the residual from the fit was close to the estimated standard deviation of the noise. Using these values, the amplitudes are recovered using a pseudo inverse operator which uses the signal and the model functions, $B = M^{\dagger}S(t)$. The 1D histogram in the main manuscript considers only the modes with frequencies above 2.5 GHz and line widths greater than 0.2 GHz. This avoids analyzing signal artifacts that arise from the timing electronics, photodiode detector, amplifiers, and digitizer, whereby the signal does not decay. The 1D bins are set to 0.25 GHz. For the 2D histogram in the main manuscript, only the modes whose amplitude is greater than 0.25 times the maximum amplitude in the set are displayed for ease of visualization. The frequency bins are set to 0.5 GHz and the line width bins are set to 0.5 GHz.

**(XIIV) Non-Virus Background Signals:** Figure S17 shows signal levels from non-virus background. In general, the pump-probe signal is much larger than any of the background signals. Here we show the signal comparison between virus signal and other background signals, which are collected by positioning the virus particle outside of the laser beam focus. We also plot the signal level of paraformaldehyde when the laser beams focus onto a small flake of this crystal. The signal level of all these signals is <0.004 V, while the virus signal amplitude varies from 0.005 to 0.024 V within the initial period of 2 ns. As previously mentioned, there are some noise spikes that arise from the electronics, but these are non-decaying and easy to discriminate from the real, decaying signal contributions.

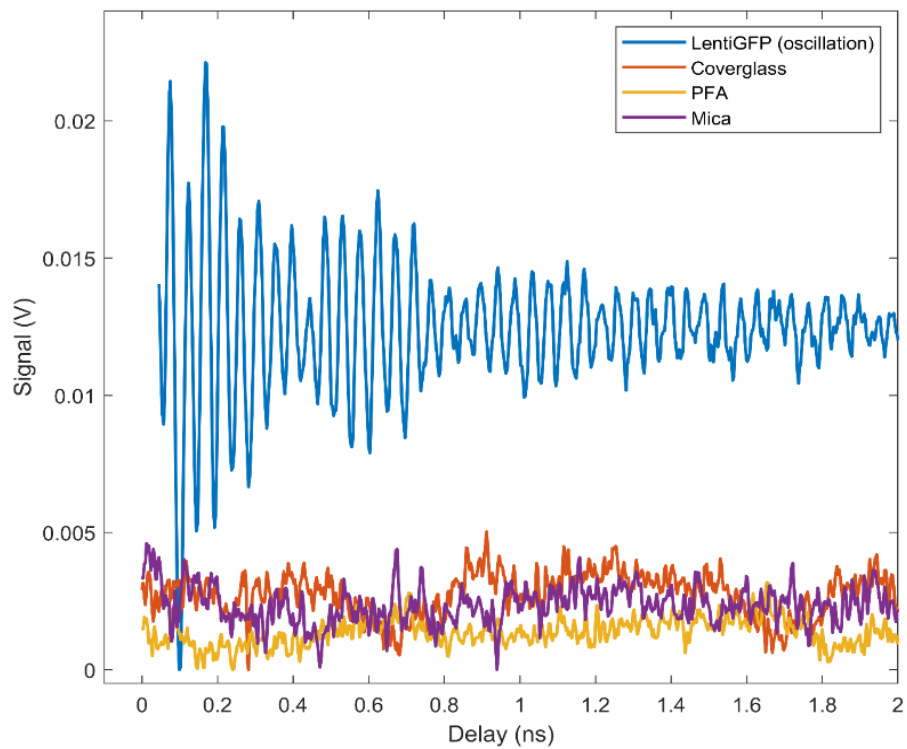

**Figure S17.** Signal level comparison between virus and other non-virus background. No large oscillations in the range 3 – 30 GHz are observed for the background (fused silica cover glass, paraformaldehyde (PFA) crystals, and mica substrate).